\documentclass[preprint]{aastex}

\begin{document}


\title{The Globular Cluster Population of NGC 7457: Clues to the
Evolution of Field S0 Galaxies} 

\author{Jonathan R. Hargis and Katherine L. Rhode} 
\affil{Indiana University, 727 East 3rd Street,
Swain West 319, Bloomington, IN 47405,USA}
\email{jhargis@astro.indiana.edu}

\author{Jay Strader\altaffilmark{1}} 
\affil{Harvard-Smithsonian Center for
Astrophysics, 60 Garden Street, Cambridge, MA 02138,USA}

\altaffiltext{1}{Hubble Fellow, now Menzel Fellow at Harvard College
  Observatory, Cambridge, MA, USA}

\author{Jean P. Brodie}
\affil{UCO/Lick Observatory, University of California, Santa Cruz, CA
  95064, USA}

\shorttitle{Globular Cluster System of NGC 7457}
\shortauthors{Hargis et al.}

\received{}
\accepted{}


\begin{abstract}
In this paper we present the results of a wide-field imaging study of
the globular cluster (GC) system of the field S0 galaxy NGC 7457. To
derive the global properties of the GC system, we obtained deep $BVR$
images with the WIYN 3.5 m telescope and Minimosaic Imager and studied
the GC population of NGC 7457 to a projected radius of $\sim 30$
kpc. Our ground-based data were combined with archival and published
{\it Hubble Space Telescope} data to probe the properties of the GC
system close to the galaxy center and reduce contamination in the GC
candidate sample from foreground stars and background galaxies.  We
performed surface photometry of NGC 7457 and compared the galaxy's
surface brightness profile with the surface density profile of the GC
system. The profiles have similar shapes in the inner $1\arcmin$ (3.9
kpc), but the GC system profile appears to flatten relative to the
galaxy light at larger radii. The GC system of NGC 7457 is noticeably
elliptical in our images; we measure $\epsilon = 0.66 \pm 0.14$ for
the GC distribution, which is consistent with our measured ellipticity
of the galaxy light. We integrated the radial surface density profile
of the GC system to derive a total number of GCs $N_{\rm GC} = 210 \pm
30$.  The GC specific frequency normalized by the galaxy luminosity
and mass are $S_N = 3.1 \pm 0.7$ and $T = 4.8 \pm 1.1$, respectively.
Comparing the derived GC system properties and other empirical data
for NGC 7457 to S0 formation scenarios suggests that this field S0
galaxy may have formed in an unequal-mass merger.
\end{abstract}

\section{Introduction}\label{intro}

As some of the longest-surviving stellar populations in galaxies,
globular clusters (GCs) provide a unique historical record of the
major star formation events and baryonic mass assembly processes in
galaxies.  For example, observations of ongoing galaxy mergers
\citep{ho92,wi93} have discovered populations of bright, blue, massive
clusters; follow-up studies indicate that these are likely proto-GCs
forming in merger and/or starburst events \citep{ho96}.  Most galaxies
contain two subpopulations of GCs as traced by their photometric
colors; bluer GCs are metal-poor and redder GCs are metal-rich for
stellar populations older than $\sim 2$ Gyr \citep{br06}.  Studies of
the two Milky Way GC populations -- the metal-poor halo GCs and
metal-rich thick disk/bulge GCs -- have provided key insights into the
formation of the Galaxy \citep{se78,zi93}, allowing for the comparison
of model predictions to the observational properties of the Galactic
GC system (see Ashman \& Zepf 1998 and references therein).  As
compact, luminous (average $M_V \sim -7$; Ashman \& Zepf 1998)
objects, GCs can be easily detected in galaxies at distances of
several tens of Mpcs.  Because photometric and spectroscopic
techniques yield estimates of GC ages, metallicities, and kinematics
\citep{br06}, studies of extragalactic GC populations can deepen and
refine our understanding of galaxy formation and evolution.

The GC systems of giant galaxies are typically more spatially extended
than the host galaxy light, often extending to many galaxy effective
radii \citep{ha91,as98,br06}.  As such, imaging studies using
detectors with small spatial coverage can only directly measure {\it
local} GC system properties.  However, the {\it global} properties of
GC systems can be accurately and directly measured from wide-field
imaging data thereby minimizing extrapolations to global
characteristics from data with less spatial coverage (see Rhode,
Windschitl, \& Young 2010 and references therein).  The determination
of the global properties of extragalactic GC systems -- including
total numbers of GCs, specific frequencies, spatial distributions,
color distributions, and color gradients -- has been the focus of our
ongoing wide-field imaging study of galaxies across a range of
environments and morphologies. These properties provide important
constraints to galaxy formation models; such models must reproduce the
observed ensemble properties of GC systems.  The survey methods and
previous results are presented in Rhode \& Zepf (2001, 2003, 2004;
hereafter RZ01, RZ03, RZ04) and Rhode et al. (2005, 2007, 2010;
hereafter R05, R07, R10).  This paper presents a study of the GC
system of NGC 7457, a lenticular (S0) galaxy included in our survey
sample (see Table~\ref{props} for basic properties).  Since NGC 7457
is a field galaxy ($\rho = 0.13 {~\rm Mpc}^{-3}$; Tully 1988), this
allowed us to examine the GC system properties of an S0 in an
environment different from most lenticulars. This environmental
difference is of primary importance when considering various S0 galaxy
formation scenarios \citep{bo06}.

Although lenticulars make up a significant fraction of the galaxy
population in nearby galaxy clusters \citep{dr97} and are important
for understanding galaxy formation and evolution, the number of
uniform and systematic studies of S0 GC systems {\it as a class} is
quite small. \citet{ku01b} used archival {\it HST}/WFPC2 data to
examine the GC systems of 29 lenticulars, providing a sample for
comparison to their {\it HST}/WFPC2 archival data study of the GC
systems of ellipticals \citep{ku01a}.  These studies found similar
mean $V-I$ colors ($V-I \sim 1-1.04$) for the GC populations in the
two galaxy classes.  Because the {\it HST} imaging was relatively
shallow, \citet{ku01b} only detected color bimodality in one S0 and
used statistical tests to estimate that $\sim 10-20\%$ of S0s are
likely bimodal.  The ACS Virgo Cluster Survey (ACSVCS; Peng et
al. 2006) imaged the central $3.4\arcmin \times 3.4\arcmin$ (17 kpc
$\times$ 17 kpc) of 32 Virgo Cluster S0s.  The deeper imaging of this
survey discovered color bimodality in $\sim 84\%$ of lenticulars.  The
ACSVCS also confirmed the mean GC color-galaxy luminosity relation in
early-type galaxies \citep{br91}; the lenticulars and ellipticals form
an overlapping sequence of redder mean GC colors for increasing galaxy
luminosity.  While these results hint at similarities between the GC
systems of early-type galaxies, at present the global properties of
the GC systems of S0 galaxies remain poorly understood relative to
giant ellipticals.

We obtained $BVR$ wide-field imaging of the S0 galaxy NGC 7457 using
the WIYN 3.5 m telescope\footnote{The WIYN Observatory is a joint
facility of the University of Wisconsin, Indiana University, Yale
University, and the National Optical Astronomy Observatory.} and
Minimosaic imager. Our ground-based imaging was complemented by the
inclusion of published and archival {\it HST} data; the high spatial
resolution data allowed us to study the GC population closer to the
galaxy center.  We also performed surface photometry of NGC 7457 in
order to compare the spatial distributions of the GC system and the
galaxy field star population.  Previous studies of the GC system of
NGC 7457 were done by Chapeleon et al. (1999; hereafter C99) and
Chomiuk, Strader, and Brodie (2008; hereafter CSB08). C99 performed
ground-based $BVI$ imaging of NGC 7457 with a similar areal coverage
($10\arcmin \times 10\arcmin$) to our WIYN study.  CSB08 used archival
{\it HST}/WFPC2 images and Keck spectroscopy to probe the ages and
metallicities of GCs in NGC 7457.

Our study of the GC system of NGC 7457 is presented as follows.  In
$\S \ref{observations}$ we describe the observations and data
reduction.  The surface photometry results are presented in $\S
\ref{sfcphotometry}$.  The methods for the detection of the GC system
are given in $\S \ref{detection}$ and the analysis of the GC system is
given in $\S \ref{analysis}$.  The global properties of the GC system
are described in $\S \ref{global}$.  A summary of this work and
conclusions are given in $\S \ref{conclusions}$.

\section{Observations and Data Reduction}\label{observations}

\begin{figure}
\plotone{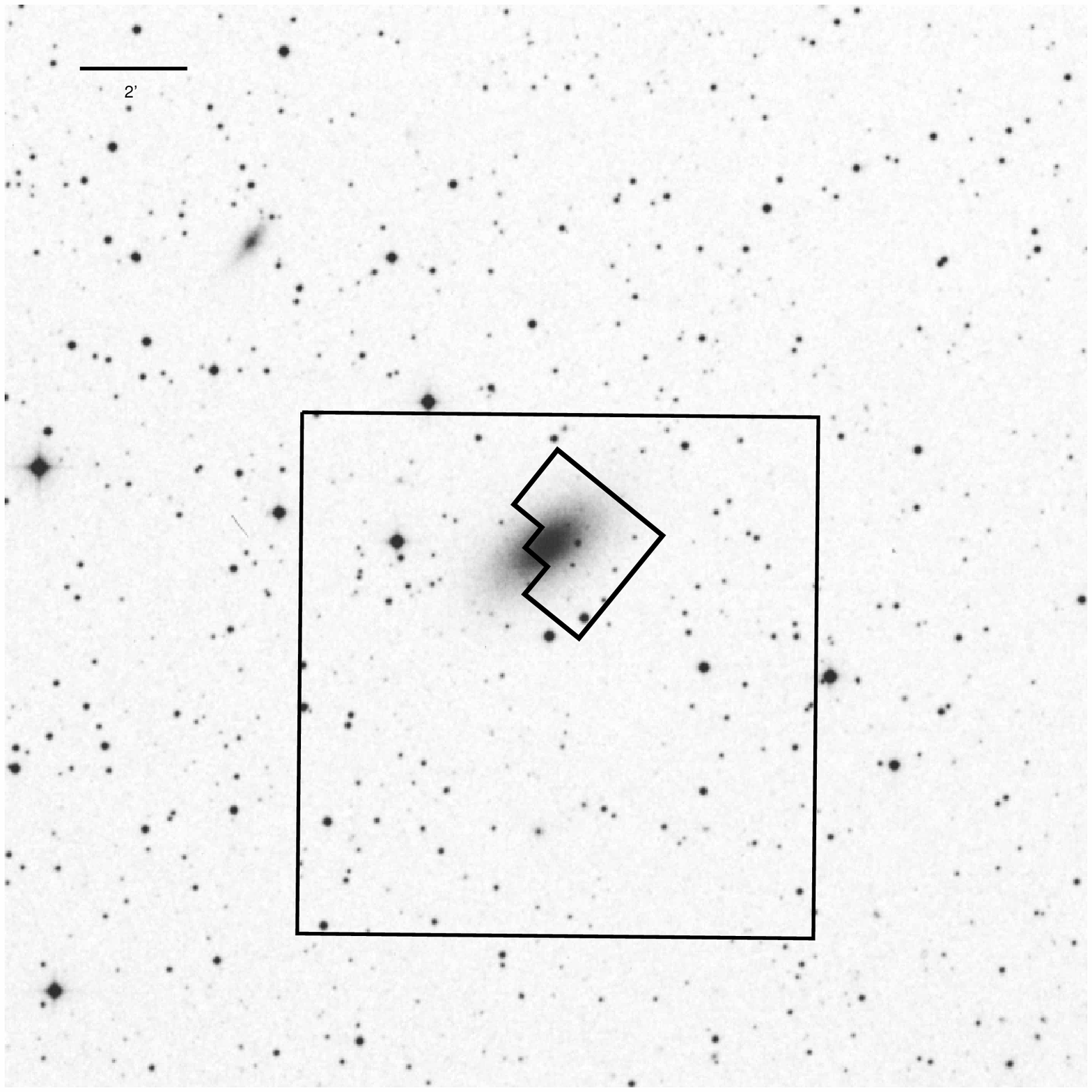}
\caption{Image of NGC 7457 from the Digitized Sky Survey showing the
  area covered by the WIYN pointings (large box) analyzed in this
  study.  The \textit{HST}/WFPC2 pointing analyzed by CSB08 is also shown.
  The orientation of the frame is north-up, east-left.
  \label{finder}}
\end{figure}

Images of NGC 7457 were obtained in broadband $BVR$ filters over three
nights in October 2009 using the Minimosaic camera on the WIYN 3.5 m
telescope at Kitt Peak National Observatory.  The Minimosaic
instrument consists of two $2048 \times 2048$ CCD detectors; on WIYN,
it has a plate scale of $0.14\arcsec$ per pixel and a field of view of
$9.6\arcmin \times 9.6\arcmin$.  For the imaging, the object was not
centered on the Minimosaic frame but was offset to increase radial
coverage around the galaxy.  Figure~\ref{finder} shows the location of
the WIYN pointing overlaid on a Digitized Sky Survey \footnote{The
Digitized Sky Surveys were produced at the Space Telescope Science
Institute under U.S. Government grant NAG W-2166. The images of these
surveys are based on photographic data obtained using the Oschin
Schmidt Telescope on Palomar Mountain and the UK Schmidt
Telescope. The plates were processed into the present compressed
digital form with the permission of these institutions.} image of the
surrounding area.  Good-quality imaging data of NGC 7457 were obtained
on nights 1 and 3 of the run; total exposure times were $3 \times 2100
\rm{~s}$ in $B$, $3 \times 2000 \rm{~s}$ in $V$, and $4 \times 1800
\rm{~s}$ in $R$.

Because the sky conditions were photometric on night two, images of
\citet{la92} standard fields and single images of NGC 7457 were
obtained in all filters. From these data we derived a
photometric calibration solution consisting of color coefficients and
zero points.  The standard errors on the zero points ranged from
$0.005$ to $0.010$.  While night two was photometric, the seeing was
worse than on nights one and three and hence these data were only used
for photometric calibration.

The images from nights one and three were used to construct one
combined, deep exposure per filter of NGC 7457 and its GC system.
First the data were overscan and bias level corrected as well as flat
fielded using standard tasks in the MSCRED package of IRAF.  Next,
single-format FITS images were constructed from the multi-extension
FITS files using the routines \texttt{msczero}, \texttt{msccmatch},
and \texttt{mscimage}.  All images in all filters were then aligned to
a chosen reference pointing. Next, for each image we measured and
subtracted a constant sky background level.  For each filter, the
corresponding images were scaled to a common flux level using the flux
measurements of 10-20 bright stars on each image.  The scale factors
were applied and images in the same filter were combined using the
\texttt{imcombine} task with \texttt{ccdclip} pixel rejection.
Lastly, the measured sky background of the image scaling reference
frames were added back to the stacked images in each filter.  The
resultant mean full-width at half-maximum (FWHM) of the image
point-spread functions (PSFs) for these final combined images are
$0.8\arcsec$ in $V$, $0.7\arcsec$ in $B$, and $0.7\arcsec$ in $R$.

\section{Surface Photometry of NGC 7457}\label{sfcphotometry}

In order to derive a total apparent magnitude for NGC 7457 and compare
the surface brightness profile of the galaxy to its GC system, we
performed surface photometry on the final, deep $V$ band image. First,
a constant sky level was removed by measuring the modal value of the
background in a $5\arcmin \times 5\arcmin$ region of the image away
from the galaxy light.  This value was in excellent agreement with the
median sky value determined for the image stacking and scaling
described in $\S \ref{observations}$. Both unresolved and extended
sources in the NGC 7457 field were masked to prevent them from
contaminating the galaxy photometry.  Surface photometry was performed
using the ELLIPSE routine \citep{je87} in IRAF. We allowed the routine
to fit ellipses from $2\arcsec$ to $200\arcsec$.  At $\sim
140\arcsec$, the ellipse solutions became unstable, indicating that at
this radius we have reached the image background level and have
insufficient signal-to-noise for reliable photometry.

\begin{figure}
\plotone{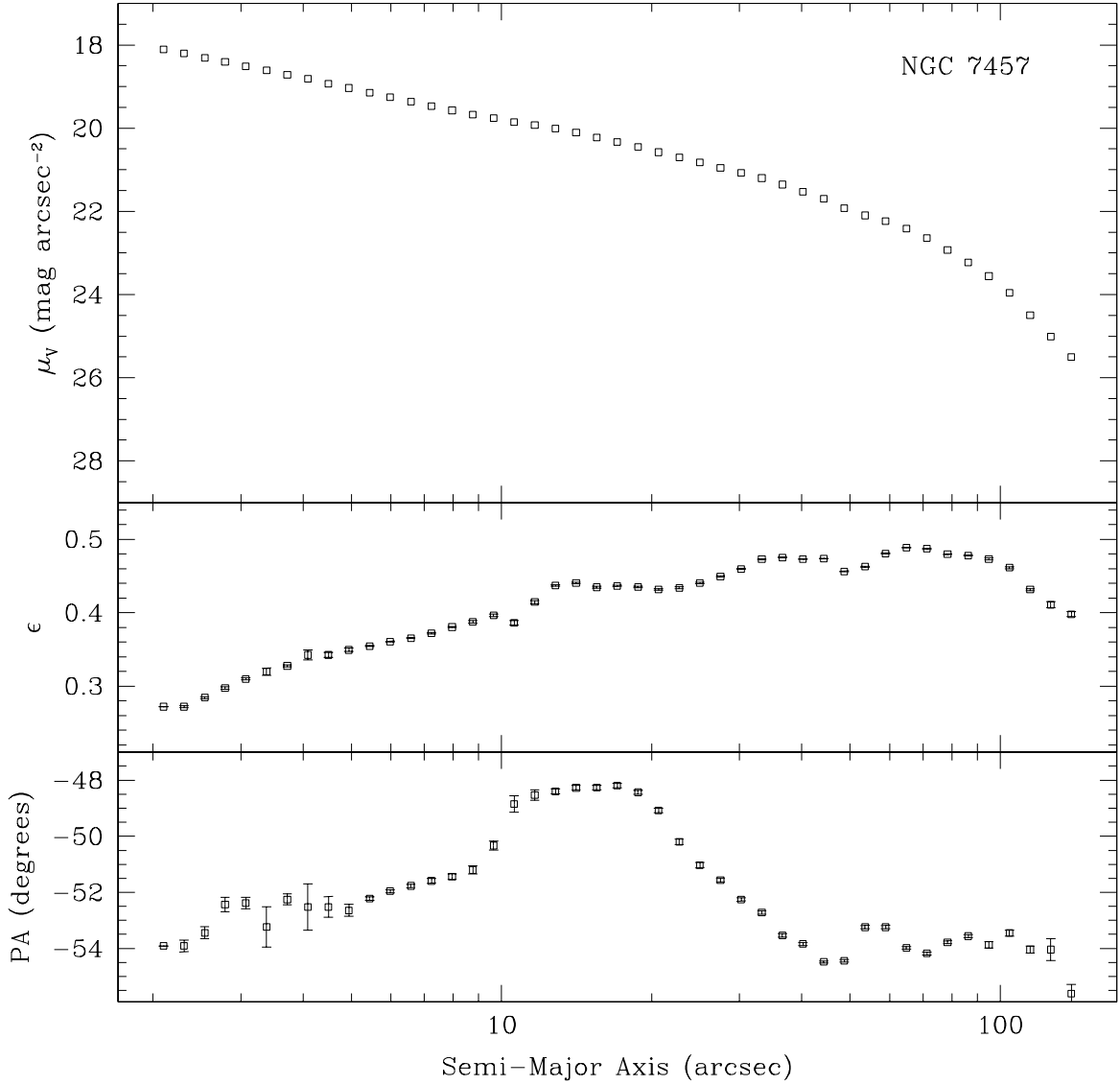}
\caption{Results of the surface photometry of NGC 7457. Shown are the
  $V$ band surface brightness profile (top panel), ellipticity (middle
  panel), and position angle (bottom panel; measured in degrees east
  of north) as a function of semi-major axis.  Error bars are shown
  for the ellipticity and position angle data; errors on the surface
  brightness are much smaller than the symbol and are not plotted.
  \label{sfc_phot_plot}}
\end{figure}

Figure~\ref{sfc_phot_plot} shows the results of the surface
photometry; the surface brightness, ellipticity, and position angle
are plotted as a function of the semi-major axis of the ellipses.  The
data are given in Table~\ref{sfc_phot_table}.  The surface brightness
profile shows a smooth decline in log-log space with an abrupt change
in the profile at $\sim 70\arcsec$; we describe the profile in more
detail below and compare the profile to the surface density of GC
candidates in $\S \ref{radial}$.  The ellipticity of the isophotes
change quite rapidly in the inner $20\arcsec$ but remains fairly
constant at an average value of $\epsilon=0.46$ from $20\arcsec$ to
$140\arcsec$. We find evidence of twisty isophotes from the variation
in position angles of the ellipses, although the range of overall
variation from $\sim -54^{\circ}$ to $\sim -48^{\circ}$ is quite
small.  From interpolation within the surface brightness profile, we
find the semi-major axis where the surface brightness reaches $\mu_V =
25~{\rm mag ~arcsec}^{-2}$ is $r_{25}=2.1\arcmin$ ($8.1$ kpc for
our assumed distance; Table~\ref{props}), in excellent agreement with
the $B$-band RC3 value of $D_{25}/2=2.1\arcmin$ \citep{RC3}.  From the
interpolated ellipticity at this radius we find a minor-to-major axis
ratio at $r_{25}$ of $0.60$, in good agreement with the RC3 value of
$0.53$ \citep{RC3} and the Bright Galaxy Catalog (BGC) value of $0.62$
\citep{tu88}.  Following the formulation of \citet{tu88}, we calculate
an inclination of $59^{\circ}$, in good agreement with BGC value of
$56^{\circ}$.

The shape of the surface brightness profile for NGC 7457 was
investigated by fitting the data with a S\'ersic profile of the form 

\begin{equation}
I = I_o {e^{-(r/r_o)}}^{\beta}, \label{sersic}
\end{equation}

\noindent where $I$ indicates the surface brightness in units of $V$
band luminosity per square parsec and the traditional S\'ersic index
$n$ is given by $1/\beta$.  Thus $\beta=0.25$ corresponds to a de
Vaucouleurs profile and $\beta=1$ corresponds to an exponential
profile.  In magnitudes per square arcsecond, this profile has the
form

\begin{equation}
\mu = {\mu}_o + 1.0857 \left(\frac{r}{r_o}\right)^{\beta}. \label{final}
\end{equation}

A fit to the entire surface brightness profile yields a S\'ersic index
of $n=2.1 \pm 0.10$ ($\beta=0.47 \pm 0.02$) but with a large $\chi^2$
value. Restricting the fit to the inner region of the profile shows
that the surface brightness follows closely to a de Vaucouleurs law
with $n=3.4-4.3$ ($\beta=0.23-0.29$), depending on the radial extent
considered as the inner region.  This profile generally describes fits
ranging from $0\arcsec$ to $20\arcsec$ and from $0\arcsec$ to
$80\arcsec$.  In the outer regions of the galaxy, the light profile is
better described by an exponential decline; we find a range of
S\'ersic indices from $n=0.98-1.02$, again depending on the choice of
where the outer profile begins. 

Our exploration of the S\'ersic fits over various ranges indicate the
that surface brightness profile is a combination of a de Vaucouleurs
law in the inner regions with an exponential decline in the outermost
regions.  This is consistent with an expectation of a bulge and disk
component for this S0 galaxy.  Fisher \& Drory (2008; 2010) used
bulge/disk decomposition profile fitting to study the optical and
infrared surface brightness profiles of NGC 7457.  They found bulge
S\'ersic indices of $n_b = 2.4 \pm 0.7$ ($V$ band; Fisher \& Drory
2008) and $n_b = 2.7 \pm 0.4$ ($3.6 \mu$; Fisher \& Drory
2010). Although these values are close to the $n_b=2$ boundary between
pseudobulges ($n_b \lesssim 2$) and classical bulges ($n_b \gtrsim 2$;
Fisher \& Drory 2008, 2010), this suggests the presence of a
featureless, elliptically-shaped classical bulge rather than a
flattened, disky-shaped pseudobulge. We discuss the implications of
NGC 7457 having a possible classical bulge in $\S \ref{conclusions}$.

The total apparent magnitude of NGC 7457 was computed according to the
following procedure.  First, the integrated magnitude of the galaxy to
the outermost elliptical annulus was computed by the ELLIPSE routine
as part of the surface photometry.  To extrapolate to an infinite
radius and obtain the total magnitude, we fit an exponential profile
to the outer portion of the surface brightness profile.  An
integration of this profile from the outermost elliptical annulus to
infinity gave an additional $0.05$ magnitudes in $V$ band.  The
derived total apparent magnitude is $V_T=11.24$, in good agreement
with the RC3 value of $V_T=11.20$ \citep{RC3}.  Assuming no internal
galaxy extinction and a negligible k-correction, applying a Galactic
extinction correction of $A_V=0.168$ (Table~\ref{props}) gives a
total, extinction-corrected magnitude of $V_T^0=11.07$.  The derived
value is between the recent measurement by \citet{fi08} of
$V_T^0=11.27$ and the RC3 value of $V_T^0=10.93$.  For the remainder
of this study, we adopt an absolute magnitude of $M_V=-19.54$ from the
derived total apparent magnitude and assumed distance modulus of
$(m-M)_V=30.61$ \citep{to01}.

\section{Detection of the Globular Cluster System}\label{detection}

\subsection{Point Source Detection and Aperture Photometry}\label{ptsource}

The detection of the GC system of NGC 7457 was performed following the
procedure developed from our previous wide-field survey studies (RZ01;
RZ03; RZ04; R07; R10).  We begin by smoothing each final combined
image with a circular median filter with diameter $= 7.5 \times
\rm{~FWHM}$. The smoothed image is then subtracted from the original
and the sky background level is added back to the frame.  All
remaining steps in the analysis were performed on the
galaxy-subtracted images. All sources on the image were detected using
the DAOFIND routine in IRAF with a $5\sigma$ detection threshold for
the $VR$ images and a $4.5\sigma$ threshold for the $B$ image. Regions
with high noise levels and/or saturated pixels, such as along the
frame edges, near bright stars, and in the saturated center of the
galaxy, were masked out. A total of 1050 objects were detected in all
three filters.

At the distance of NGC 7457 and given the resolution of the images
($\sim 1\arcsec$), we expect the GC population to appear as point
sources in the frames.  We used a graphical software routine to refine
our list of detected objects by removing extended sources. Point
sources were isolated by plotting the FWHM of the objects versus
instrumental magnitude, selecting objects around the well-defined
sequence delineated by the points sources.  The scatter in the point
source sequence increases at fainter magnitudes; in this regime we
allowed for a larger range of selected objects.  After requiring that
objects appear as point sources in all three filters we are left with
540 objects.

For the point source detections we performed aperture photometry in
order to obtain calibrated magnitudes. Aperture corrections were
calculated by measuring the mean difference between the total
magnitude and the magnitude within an aperture of radius $1\times{\rm
FWHM}$.  The aperture corrections ranged from -0.196 to -0.294 with
errors of 0.001-0.003.  Because the photometric solution was derived
from data taken on night two and the final $BVR$ science images were
constructed using data from nights one and three (see $\S
\ref{observations}$), bootstrap offsets to the instrumental magnitudes
of night two were determined using the final, stacked science images
and the single images of NGC 7457 taken on night two.  These offsets
range from -0.061 to -0.063 with errors of 0.003 to 0.007.  The final,
calibrated total magnitudes for image point sources were calculated by
applying the aperture corrections, atmospheric extinction correction,
bootstrap offsets, and photometric calibration coefficients to
instrumental magnitudes derived from a $1\times{\rm FWHM}$ radius
photometric aperture.  Lastly, Galactic extinction corrections of
$A_B=0.219$, $A_V=0.168$, and $A_R=0.136$ magnitudes were applied to
the calibrated data; the corrections were obtained from the reddening
maps of \citet{sc98}.  The effect of an uncertainty on the extinction
corrections on the GC candidate selection is considered in $\S
\ref{colorcut}$.

\subsection{Color Selection}\label{colorcut}

The final step in deriving a list of GC candidates is to choose
objects with $BVR$ colors and magnitudes that are consistent with
those of nearby, confirmed GCs at the distance of NGC 7457.  This
method of selecting GC candidates has been used in our previous survey
studies (RZ01; RZ03; RZ04; R07; R10).  Here we outline the methodology
that is described in detail in RZ01.  Assuming a range of GC absolute
magnitudes from $M_V \sim -11$ to $-4$ based on the globular cluster
luminosity function (GCLF) of the Milky Way and other well-studied
nearby galaxies \citep{as98}, we eliminate point sources that are
brighter than the brightest GC we would expect at the distance of NGC
7457. We select GC candidates with $B-V$ and $V-R$ colors consistent
with those of the Milky Way GCs, but allow for a larger range of
metallicities ([Fe/H] = -2.5 to 0.0).  Because low luminosity galaxies
have smaller total numbers of GCs, we carefully consider those objects
which may just barely fail the selection criteria to make sure we have
included all likely GC candidates.  Additional data by which we can
judge the probability of a point source being considered a GC are the
radial proximity to the galaxy center and the appearance of the object
in archival $HST$ images (see $\S \ref{background}$).

\begin{figure}
\plotone{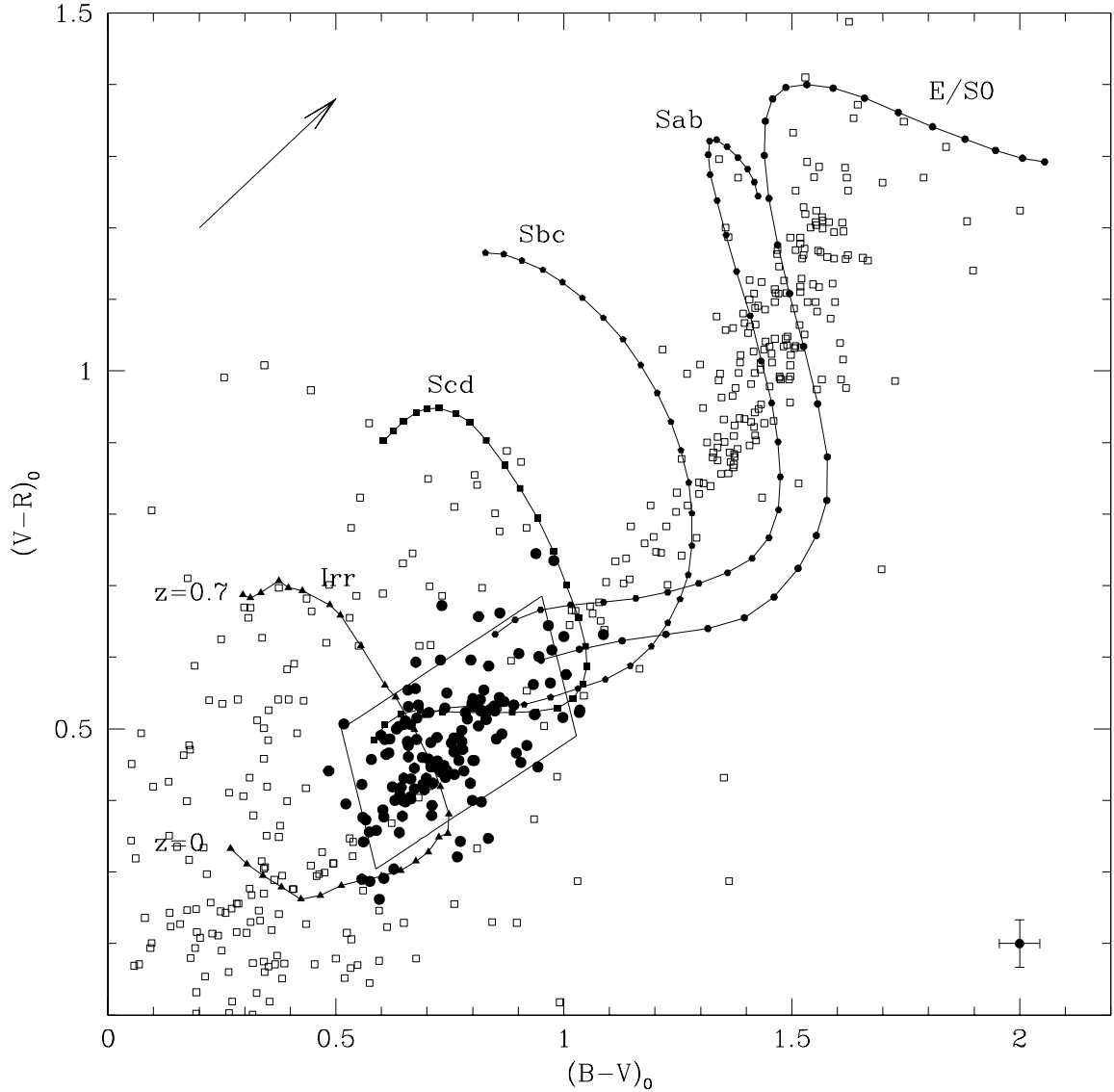}
\caption{Selection of GC candidates in NGC 7457 based on $B-V$ and
  $V-R$ colors and $V$ magnitudes.  The 540 point sources detected in
  all three filters are shown as open squares.  The final sample of
  136 GC candidates is shown as filled circles.  The box shows the
  boundary of our color selection (see $\S \ref{colorcut}$). The
  reddening vector corresponding to $A_V=1$ is shown in the upper
  left-hand corner.  The curves illustrate the location of galaxies of
  various galaxy morphological types from redshift $z=0$ to $0.7$.
  The lower right corner shows the median $V-R$ and $B-V$ photometric
  errors for the WIYN GC candidates.  Our imposed $V$ magnitude cut
  rejects some objects within the color selection box.  The
  consideration of photometric errors in our GC selection criteria
  causes some objects outside the color selection box to be included
  as GC candidates.
  \label{color_color_plot}}
\end{figure}

\begin{figure}
\plotone{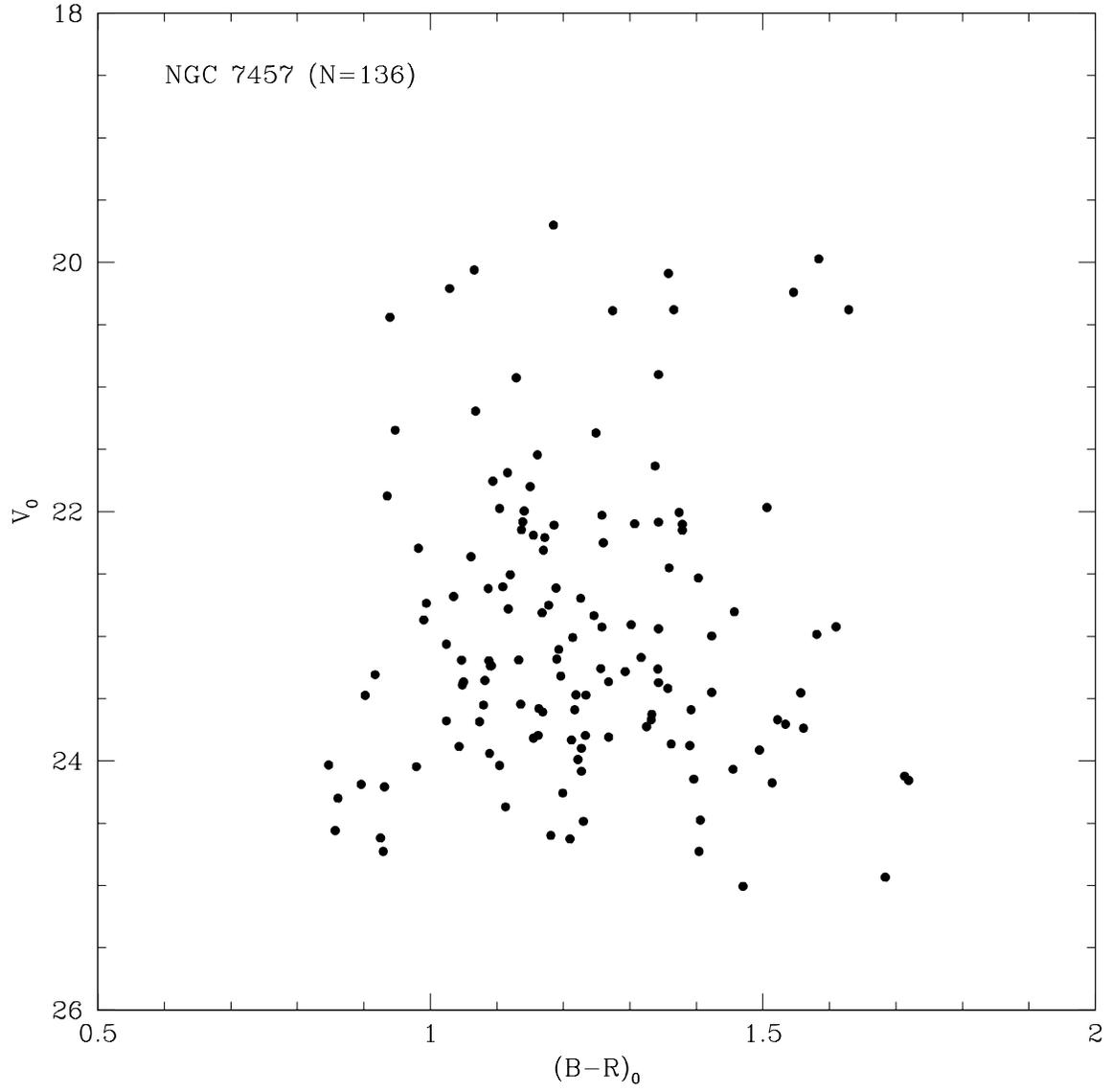}
\caption{$V,(B-R)$ color magnitude diagram for the final set of GC
  candidates in NGC 7457.  The magnitudes and colors have been corrected
  for Galactic extinction. 
  \label{cmd_plot}}
\end{figure}

Figure~\ref{color_color_plot} shows the color selection process for
the point sources that passed the extended source cut (see $\S
\ref{ptsource}$).  Also shown are the colors expected for galaxies of
various morphological types from redshifts $z=0-0.7$ (RZ01),
illustrating where contaminating background galaxies are likely to lie
in the color-color plane.  We accepted any sources with $V \geq 19.6$
and $B-V$ and $V-R$ colors (including associated photometric errors)
within $3\sigma$ of the Galactic GC $V-R$ versus $B-V$ relation. We
evaluated each object that passed the magnitude and color selection
step to make sure it should remain in the final sample. Two objects
that were barely outside the color selection but are close to the
galaxy center were added back to the GC candidate sample.  Three
objects which appeared in {\it HST} images to be background galaxies
(see $\S \ref{background}$) were removed from the GC candidate list.
The final sample consists of 136 GC candidates in the WIYN images of
NGC 7457; the color-magnitude diagram (CMD) of the GC candidates is
shown in Figure~\ref{cmd_plot}.  Because the Galactic extinction
correction is applied systematically to every object, the impact of an
uncertainty in the extinction on the GC candidate selection
corresponds to a systematic shift of all the point sources in the
color-color plane along the reddening vector shown in
Figure~\ref{color_color_plot}. The GC selection box (as defined by the
Milky Way GC system) remains unchanged. We find that significant
systematic increases or decreases in the \citet{sc98} $E(B-V)$
reddening estimate (such as doubling or halving the reddening) make a
negligible impact on our selection of GC candidates. At most only 1-2
objects would be added or removed from our GC candidate list based on
these changes.

\section{Analysis of the Globular Cluster System}\label{analysis}

\subsection{Completeness Testing and Detection Limits}\label{complete}

\begin{figure}
\plotone{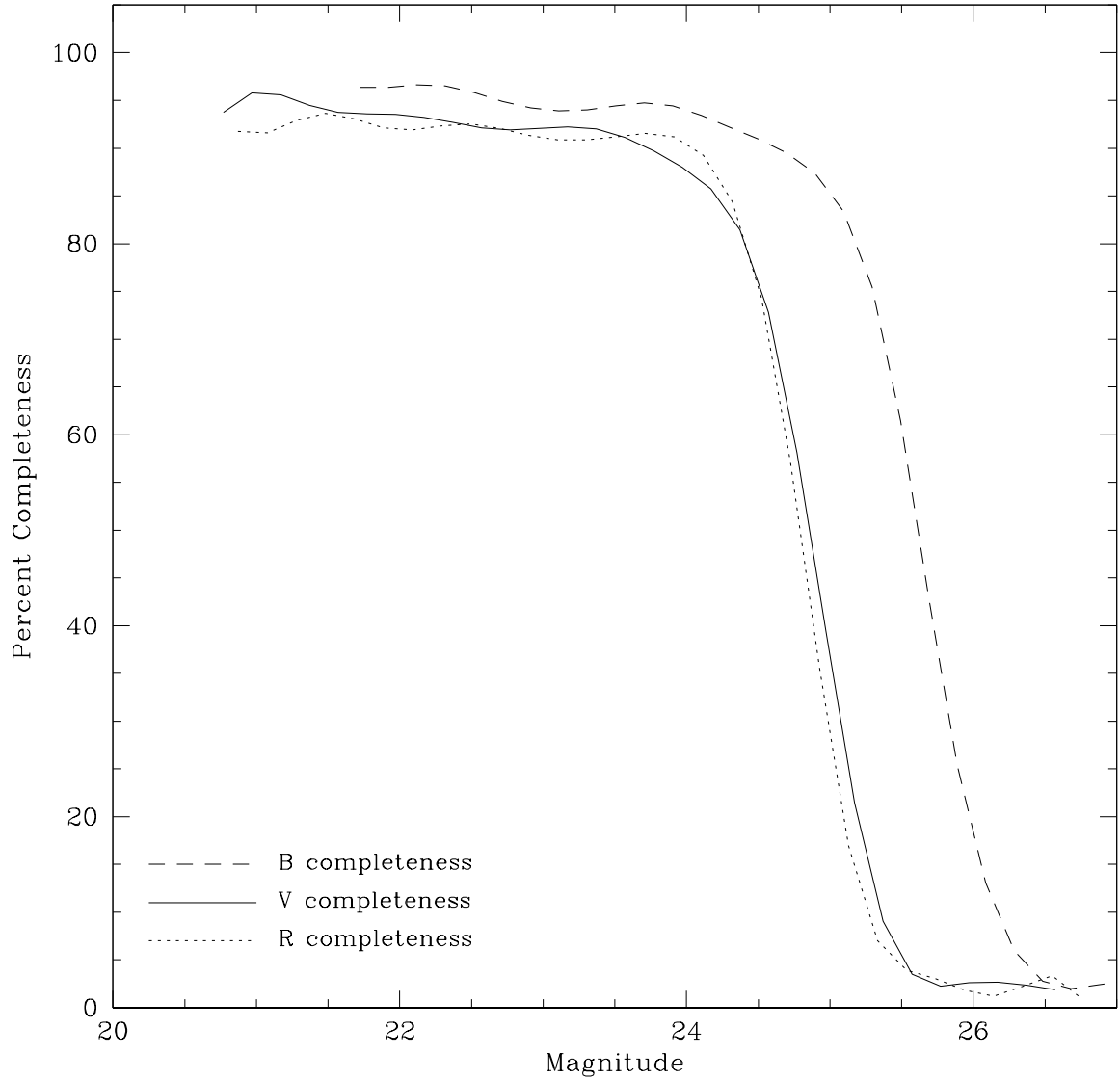}
\caption{Completeness as a function of magnitude.  The point-source
detection limits in the WIYN $BVR$ images were determined using the
artificial star test described in $\S\ref{complete}$.  The results of
the artificial star tests are shown here as completeness curves.
\label{completeness}}
\end{figure}

To quantify the point-source detection limits we performed a series of
completeness tests using the $BVR$ images employed to detect the GC
system.  For each test, 200 artificial stars were added to the image;
the magnitude of the artificial stars were chosen around 0.2 of a
specified value and the adopted PSF was taken as the best-fit average
PSF of the image.  The identical steps used to detect the point
sources in the original images were then performed, recording the
fraction of recovered artificial stars.  We repeated this process in
0.2 mag intervals over a range of 4-5 magnitudes per filter.  Taking
the fraction of recovered stars in each magnitude interval then gave a
measure of completeness as a function of magnitude which we show in
Figure~\ref{completeness}.  The WIYN images of NGC 7457 have 50\%
completeness limits of $B=25.6, V=24.8$, and $R=24.8$.

\subsection{Contamination Corrections}\label{contamination}

In ground-based imaging studies of extragalactic GC populations,
sample contamination from foreground Galactic stars and background
galaxies is a serious concern.  Following the steps from our previous
imaging studies, in $\S \ref{foreground}$ and $\S \ref{background}$ we
performed a series of analyses to help quantify the degree of
contamination from these two sources.  We ultimately adopted the
contamination estimate from the asymptotic behavior of the GC radial
profile as discussed in $\S \ref{asymp}$.

\subsubsection{Models of Foreground Galactic Contamination}\label{foreground}

To estimate the degree of contamination from Galactic foreground
stars, we used a Galactic structure model \citep{me96,me00} to predict
the surface density of stars within a given magnitude range and color.
We adopted the same Galaxy structure parameters used in our previous
studies (position of the Sun; fraction of stars in the disk, thick
disk, and halo).  Given the $V$ magnitude and $B-V$ color ranges from
the GC candidate sample, the model predicted a surface density of
stars of $0.20 {\rm ~arcmin}^{-2}$.  NGC 7457 does not have an
appreciably high Galactic latitude ($b=-27$, NED); hence this modest
degree of contamination does appear to be consistent with the large
number of point sources covering a wide range of colors in
Figure~\ref{color_color_plot}.  Because the Galactic structure models
do not constrain the range of $V-R$ colors, the model prediction is an
upper limit to the degree of Galactic foreground star contamination.

\subsubsection{\textit{HST} imaging of WIYN GC Candidates}\label{background}

An estimate of the contamination in the GC sample due to unresolved
background galaxies was obtained by examining the GC candidates in
archival \textit{HST} images.  Since many background galaxies will be
unresolved in our ground-based images, \textit{HST} imaging of point
sources in our field can show if any GC candidates are actually
background galaxies.

For NGC 7457, we downloaded \textit{HST} WFPC2 data from the
Multimission Archive at the Space Telescope Science Institute
(MAST)\footnote{Based on observations made with the NASA/ESA
\textit{Hubble Space Telescope}, obtained from the data archive at the
Space Telescope Science Institute.  STScI is operated by AURA, under
NASA contract NAS 5-26555.}.  The retrieved images were six
pipeline-reduced, multi-extension FITS files (total exposure time 1380
s) of a central pointing of NGC 7457 in the F555 filter (program
GO.5512; PI Faber).  These images were also used by CSB08 in their
analysis of the NGC 7457 GC system (see $\S \ref{hst}$).  The images
were stacked using the STSDAS task \texttt{combine}.  Thirty-three of
the WIYN GC candidates were found on the \textit{HST} image.
Following \citet{ku99}, we performed aperture photometry of these
objects in 0.5 and 3 pixel apertures.  The ratio of counts in the
larger aperture to the counts in the smaller aperture provides a
measure of the central concentration of these objects.  Slightly
extended background galaxies would show a higher count ratio than the
compact GCs, thereby providing an estimate of the fraction of the WIYN
GC candidates which might be background galaxies.  Three objects in
the WIYN GC sample were faint and had high count ratios (ratio $> 8$),
indicating that they are likely background galaxies.  In addition,
these objects did not pass the GC candidate selection criteria
employed by CSB08 ($V-I$ color cut and estimated absolute sizes).  As
noted in $\S \ref{colorcut}$, these objects were removed from the
final sample of WIYN GC candidates.  Considering these three high
count ratio objects as background galaxies and the areal coverage of
the {\it HST}/WFPC2 field, we found $0.57 \pm 0.33 ~\rm{arcmin}^{-2}$
(assuming Poisson statistics) as our estimate of the background galaxy
surface density.

Given the small areal coverage of the \textit{HST} images and the
correspondingly small numbers of GC candidates considered, our final
estimate of the surface density of contaminating objects was
determined from the asymptotic behavior of the radial profile of the
GC system.  We discuss this method and the results in $\S
\ref{asymp}$.

\subsubsection{Contamination Estimates based on the Asymptotic
  Behavior of the Radial Profile}\label{asymp}

Typical wide-field extragalactic GC studies show a central peak in a
plot of the azimuthally-averaged surface density of GC candidates as a
function of projected radial distance from the host galaxy center.
Assuming the observations have covered a sufficient area around the
galaxy, this profile will decrease to a constant level when the full
radial extent of the GC system has been measured.  Thus the outer
regions of this profile can yield a measure of the surface density of
foreground stars $+$ background galaxies (Harris 1986; R10).  A
measurement of this surface density can then be subtracted from the
radial profile to correct for contamination.

We constructed the initial radial profile of NGC~7457's GC system by
counting the number of WIYN GC candidates in concentric circular
annuli and computing the effective area in each annulus.  We explored
the shape of the radial profile in detail, examining the effects of
using wider radial bins and different locations for the bin centers.
Ultimately we adopted radial bins from $0.24\arcmin$ from the galaxy
center (inner edge of innermost annulus) to $7.74\arcmin$ (outer edge
of outermost annulus) with a width of $0.5\arcmin$.  Inward of
$0.24\arcmin$ from the galaxy center we have no GC candidates and very
little observed area due to the masking of the image.  The effective
area of each annulus is computed by taking the total area of the
annulus and subtracting the area of any masked regions in that
annulus.  The profile shows a decreasing surface density which falls
to a nearly constant level of $0.54 \pm 0.20{\rm ~arcmin}^{-2}$ in the
outer three radial bins ($\sim 6-7\arcmin$), indicating that we
have observed the full outer radial extent of the GC system.

The constant level in the three outer bins of the uncorrected radial
profile is a good estimate of the surface density of contaminants in
the GC candidate sample. The surface density of foreground stars from
the Galactic model was $0.20{\rm ~arcmin}^{-2}$; this is smaller than
the asymptotic contaminant level but makes sense if one assumes that
background galaxies will also contribute.  The background galaxy
contamination level as estimated from the {\it HST} data was $0.57
{\rm ~arcmin}^{-2}$, which is higher than the level from the
asymptotic behavior of the profile.  It is quite possible that the
former is an overestimate and it also has a large uncertainty.  Thus
we adopted the observed asymptotic correction as the final estimate of
the surface density of contaminants.  We computed the contamination
fraction for each annulus of the radial profile by multiplying the
surface density of contaminating objects by the effective area of the
annulus and then dividing this number by the number of GC candidates
in that bin.

\subsection{Coverage of the GCLF}\label{gclf}

\begin{figure}
\plotone{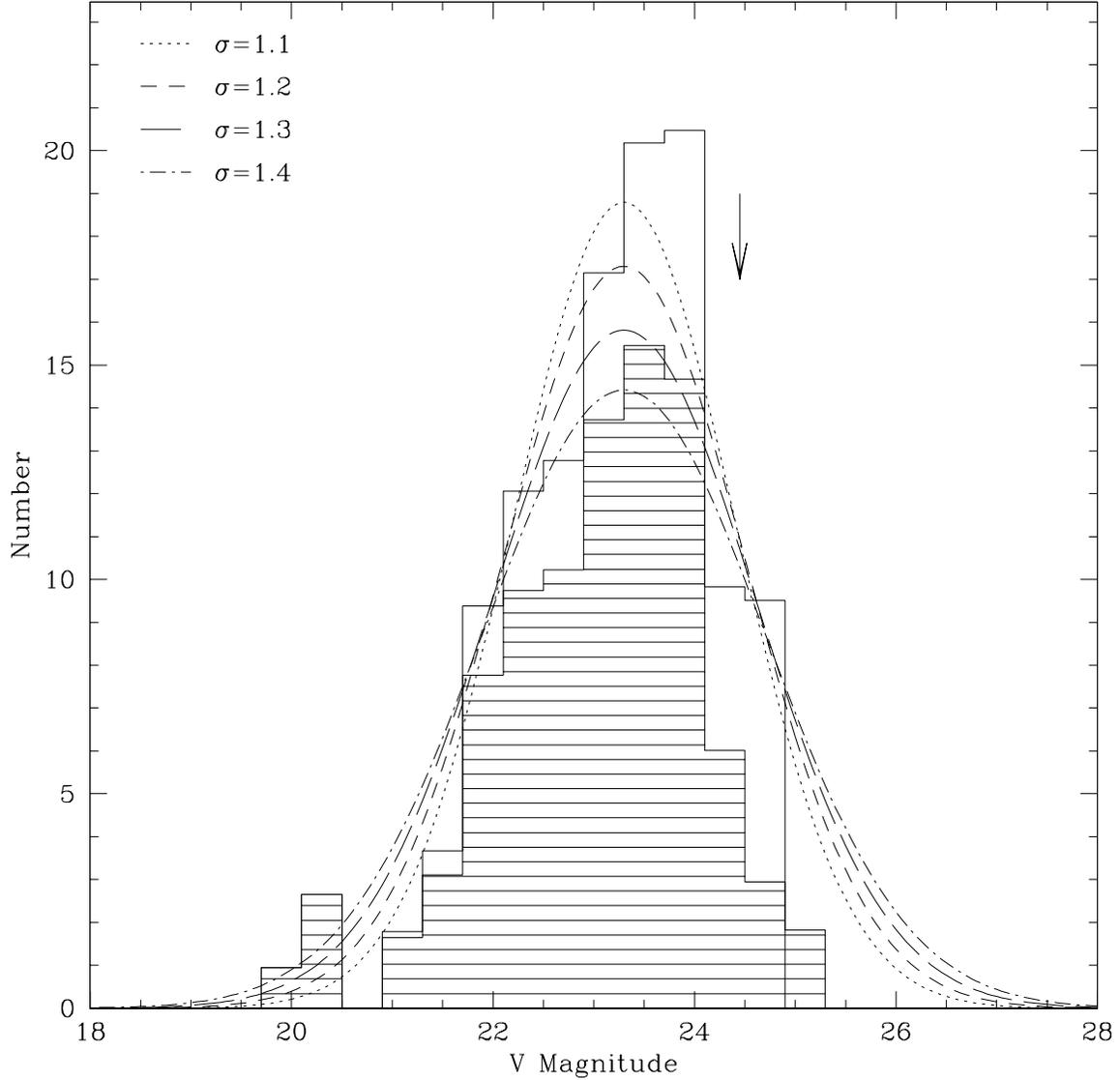}
\caption{GCLF fitting for NGC 7457.  The shaded histogram shows the
observed (contamination-corrected) GC luminosity function.  The solid
line histogram shows the completeness-corrected GC luminosity function
used in the fitting (see $\S\ref{gclf}$).  The faintest three bins
were not used in the fitting.  The best-fit theoretical Gaussian
functions are shown for the four values of the dispersion $\sigma$
considered in the fitting: $\sigma=1.1$ (dotted line), 1.2 (short
dashed line), 1.3 (long dash), and 1.4 (dot-dashed line).  The arrow
shows the magnitude where the combined $BVR$ completeness is 50\%.
\label{gclf_plots}}
\end{figure}

In order to determine the fractional coverage of the GCLF, we assigned
our WIYN GC candidates to $V$ magnitude bins with a width of $0.4$
magnitudes.  We corrected the observed GCLF for both contamination and
magnitude incompleteness following the method outlined in RZ01.  We
accounted for contamination by considering the radial location of each
GC candidate, applying the radially dependent contamination fraction
determined in $\S \ref{asymp}$ as a correction factor.  Magnitude
incompleteness was determined from the results of the artificial star
tests (see $\S \ref{complete}$ and Figure~\ref{completeness}) and the
range of $B-V$ and $V-R$ colors of the GC candidates.  The
completeness corrections we applied (the total completeness) accounted
for the individual incompleteness in all three filters as described in
RZ01.  These calculated total completeness corrections were then
divided by the number of GC candidates in each magnitude bin to obtain
a corrected GCLF.  Figure~\ref{gclf_plots} shows the observed GCLF
(contamination corrected; shaded histogram) and the
completeness-corrected histogram (solid line histogram).

The corrected GCLF was then fit with a Gaussian function to determine
the observed coverage of this theoretical GCLF.  We assumed a peak
absolute magnitude of $M_V=-7.3$ based on observations of the Milky
Way GC system \citep{as98}.  At the assumed distance of NGC 7457
(Table~\ref{props}), the apparent $V$ magnitude peak of the GCLF
corresponds to $V=23.3$.  We eliminated from the fitting the faintest
three magnitude bins which had $<60\%$ completeness values. We varied
the dispersion of the Gaussian fits between $\sigma=1.1$ and $1.4$ and
found the mean fractional coverage of the theoretical GCLF by the
observed GCLF was $0.693 \pm 0.006$. The Gaussian fits are shown in
Figure~\ref{gclf_plots}.  Varying the magnitude bin sizes between
$0.3$ and $0.5$ mag yielded a change in the fractional coverage of
$8\%$ to $13\%$.  We account for this variation in our error estimates
on the total number of GCs in $\S \ref{numbers}$.

\subsection{Combining \textit{HST} Observations with the WIYN Results}\label{hst}

CSB08 used \textit{HST}/WFPC2 archival images of NGC 7457 in the F555W
and F814W filters to select a sample of GC candidates.  These
observations (proposal GO.5512; PI Faber) were taken during a single
pointing, with the PC chip centered on the galaxy center (see
Figure~\ref{finder}).  The spatial coverage of WFPC2 chip gives radial
coverage of the inner $\sim 2\arcmin$ of NGC 7457, complementing our
WIYN observations which only provide reliable data for $r \gtrsim
2.5\arcmin$. CSB08 select GC candidates based on two criteria: a $V-I$
color cut and an absolute size cut; their final GC candidate list
consisted of 77 objects.  As noted previously, CSB08 performed
follow-up spectroscopy for some of the objects, providing confirmation
as GCs for 13 of their GC candidates.

We compared the WIYN and {\it HST} GC candidate lists and found 29 of
the 77 objects are common to both samples.  Of the 48 \textit{HST} GC
candidates which were not detected as GC candidates by our WIYN study,
37 were not detected during the \texttt{daofind} step (see
$\S\ref{ptsource}$) due to their spatial proximity to saturation bleed
trails and/or their location near the galaxy center.  Only six of
these 37 objects were missed to due their magnitude being fainter than
the detection threshold in the images.  Of the remaining 11 of 48
objects that appeared in the HST list but not the WIYN list, seven
were rejected because they appeared extended in one or more of the
WIYN images (see $\S\ref{ptsource}$).  Only two of these seven
rejected objects were ``borderline'' cases. Relaxing our extended
source cut criteria would also increase contamination, so we decided
not revise our extended source cut based on these results.  Lastly, we
found that four \textit{HST} GC candidates were rejected as WIYN GC
candidates at the color cut stage (see $\S\ref{colorcut}$).  These
objects had $B-V$ and $V-R$ colors in the WIYN data that were
significantly outside our color-color selection criteria; no
reasonable modification to our color cut would have included these
objects in our WIYN sample.  An additional check on our GC selection
criteria comes from considering the thirteen
spectroscopically-confirmed GCs.  Of these thirteen objects, eight
were detected in our WIYN data; the remaining five objects fell in
masked regions of our image.  For these eight detections, we note that
none of the spectroscopically-confirmed GCs were rejected during our
GC selection process.

We constructed the radial surface density profile for the \textit{HST}
GC candidates in the following manner.  We accounted for magnitude
incompleteness by only considering the brightest {\it HST} GC
candidates, imposing a magnitude cut at the GCLF turnover magnitude of
$V \leq 23.3$.  Since contamination from background galaxies with
magnitudes $V\sim 24$ should be relatively small \citep{ku99}, we
assumed no contamination.  To maintain a uniform comparison with the
WIYN GC candidates, we imposed our color cut criteria on the
\textit{HST} GC candidate list. We removed four objects in the {\it
HST} GC sample which had $B-V$ and $V-R$ colors \textit{as measured in
the WIYN data} which were inconsistent with our color selection. After
applying these restrictions, 46 {\it HST} GC candidates were used to
construct the GCLF and radial profile. The {\it HST} GCLF was fit
according to the same procedure as the WIYN data; the mean fractional
coverage is $0.455 \pm 0.001$.  Varying the bin sizes changed the
fractional coverage by $\sim 7\%$ and we include this uncertainty in
the final error analysis.

\section{Global Properties of the Globular Cluster System}\label{global}

\subsection{Radial Distribution of the GC System}\label{radial}

\begin{figure}
\plotone{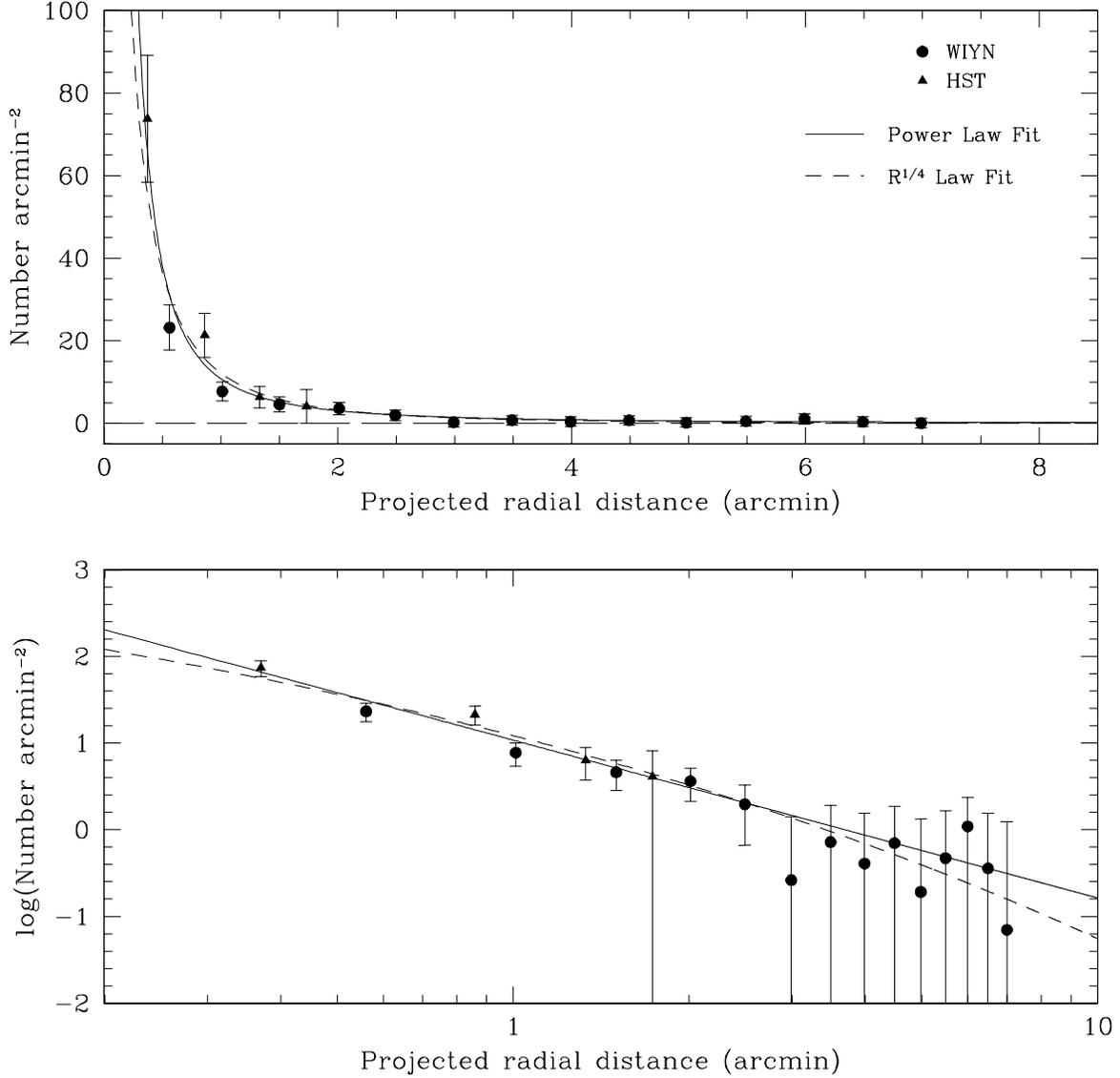}
\caption{Radial profile of the NGC 7457 GC system, corrected for
  missing area, magnitude incompleteness, and contamination (see
  $\S\ref{radial}$).  The top panel shows the surface density of GC
  candidates as a function of projected radial distance; the dashed
  horizontal line denotes a surface density of zero.  The bottom panel
  shows the logarithm of the surface density versus the projected
  radius on a logarithmic scale. The best-fit $R^{1/4}$ and power law
  profile fits discussed in $\S\ref{radial}$ are shown as the dashed
  line and solid lines, respectively.
  \label{radial_profile}}
\end{figure}

We constructed the final radial distribution of NGC~7457's GC system
by binning the WIYN and HST GC candidates into concentric circular
annuli spaced $0.5\arcmin$ apart.  The superior resolution of {\it
HST} yielded GC candidates closer to the galaxy center, so the {\it
HST} annuli begin at a projected radius $r$ of $0.08\arcmin$ whereas
the first WIYN annulus starts at $r=0.24\arcmin$.  The radial profile
was corrected for missing area, contamination (as appropriate), and
magnitude incompleteness in the following manner.  The missing area
was considered by finding the effective area for each annulus; the
area covered by the masked regions in each annulus was computed and
accounted for in the surface density calculation.  Contamination
corrections were applied to the WIYN data using the radially-dependent
contamination fractions (see $\S \ref{gclf}$); for the {\it HST} data
we assumed no contamination (see $\S \ref{hst}$).  Lastly we corrected
for magnitude incompleteness by dividing each bin by the fractional
coverage of the observed WIYN or {\it HST} GCLF, respectively.  The
surface density was computed using the effective area and corrected
total numbers for each radial bin; uncertainties in the surface
density are from Poisson errors in the number of GCs and number of
contaminating objects.  Figure~\ref{radial_profile} shows the radial
surface density as a function of projected average radius, where the
average radius is the mean radius of the unmasked pixels in that
annulus.  Table~\ref{radial_profile_table} lists the corresponding
data: average radius, surface density, errors, fractional areal
coverage of each annulus (area of unmasked pixels/total area), and
data source (either WIYN or \textit{HST}).  Because the GC system of
NGC 7457 is found to have a large ellipticity (see
$\S\ref{ellipticity}$), we experimented with the use of concentric
elliptical bins to construct the radial surface density profile.  We
found no significant dependence of the radial profile or profile fits
(see discussion below) on the choice of bin shape.

The radial profile of the NGC 7457 GC system follows the typical trend
seen in our other wide field imaging studies (RZ01; RZ03; RZ04; R07;
R10): there is a sharp rise in the surface density approaching the
galaxy center and smooth decrease with projected radius.  Although the
WIYN data does measure the surface density of GCs quite close to the
galaxy center ($r\sim 0.24 \arcmin \sim 920~\rm{pc}$), the addition of
the \textit{HST} data allows us to probe nearly three times further,
to $r\sim 0.08 \arcmin \sim 310~\rm{pc}$.  However, the
\textit{HST}/WFPC field-of-view is insufficient to cover the full
radial extent of the GC profile as evidenced by the WIYN data in the
continued decline of the radial profile outside of $r\sim 2\arcmin$.
This underscores the importance of combining by wide-field imaging to
observe the full radial extent of the surface density profile with
high-resolution \textit{HST} data which can measure the regions of the
profile closer to the galaxy center (R10).  We found that the surface
density of GCs falls to zero within the errors in the radial bin
centered at $r=2.99\arcmin$ and remains consistent with zero outward
of this radius.  R07 present an empirical relation between the radial
extent of the GC system and the galaxy mass for other galaxies
observed for our wide-field GC system survey.  The radial extent of
the GC system was defined as the radius at which the surface density
goes to zero within the errors and remains zero. This definition is
therefore sensitive to the details of our analysis, including the
magnitude completeness and the contamination correction to the radial
profile.  In addition, GCs are often found at very large radii ($>
100$ kpc for the Milky Way), so there are very likely GCs in NGC 7457
beyond our measure of the radial extent. We compared our findings on
the radial extent for NGC 7457 to the updated results presented in
R10. For our assumed distance (Table~\ref{props}), we found a radial
extent of $12\pm 2$ kpc, where the uncertainty in the radial extent
includes the uncertainty in the distance modulus and a one bin-width
uncertainty in the determination of the radial extent from the surface
density profile.  For the galaxy mass, we used a mass-to-light ratio
of $(M/L)_V=7.6$ for S0 galaxies from \citet{ze93} in combination with
our derived $M_V$ (Table~\ref{props}) and find $\log(M/M_\odot)=10.6$.
Refitting the radial extent-mass relation with a second-order
polynomial using the result for NGC 7457 with the data in R10 yields
coefficients consistent with those of R10 within the errors.

The radial distribution of GCs was investigated further by fitting the
corrected radial profile with various functional forms: a power law
profile ($\log{\sigma_{\rm GC}}=a_0 + a_1\log{r}$), de Vaucouleurs
profile ($\log{\sigma_{\rm GC}}=a_0 + a_1 r^{1/4}$), and S\'ersic
profile ($\log{\sigma_{\rm GC}}=a_0 + a_1 r^{1/n}$).  The S\'ersic
profile fitting yielded highly uncertain results; a wide range of
possible parameters give very similar $\chi^2$ values. Test data sets
with comparable numbers of data points and scatter to our observed
profile were constructed to better understand this behavior.  We found
that the small numbers of data points in inner regions of the radial
profile and a strong sensitivity to the initial guesses for the fit
parameters yield a model that is not well constrained by the data. The
de Vaucouleurs law and power law both provided good fits to the data,
intersecting nearly all the points in the profile; the reduced
chi-squared value for the power law (0.70) was slightly smaller than
that for the de~Vaucouleurs law (0.92). The fit parameters are given
in Table~\ref{profile_fits} and Figure~\ref{radial_profile} shows the
de Vaucouleurs and power law fits to the data.  We discuss the
calculation of the total number of GCs from the integration of the
radial profile in $\S \ref{numbers}$.

\begin{figure}
\plotone{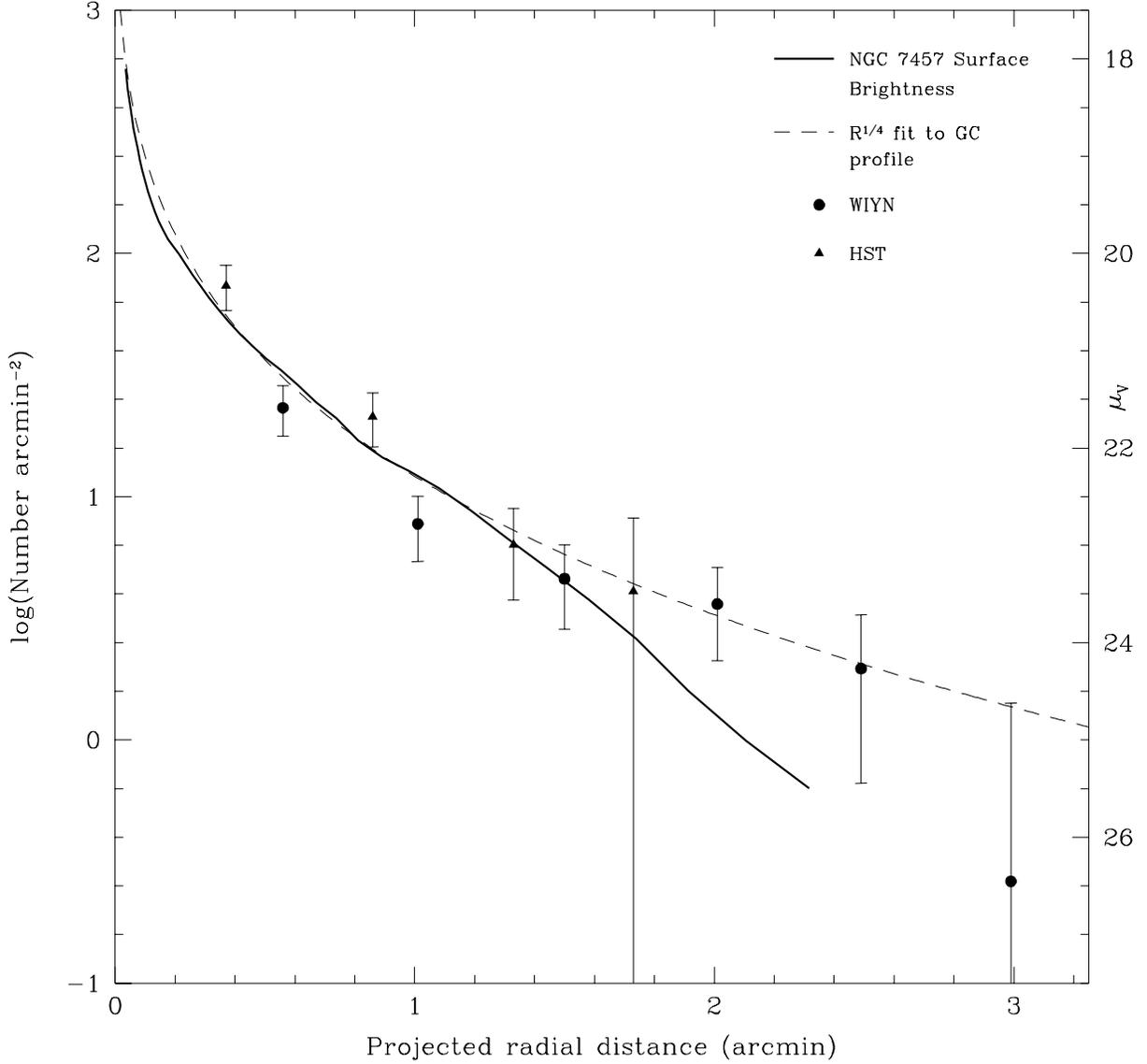}
\caption{Comparison of GC surface density and NGC 7457 surface
  brightness profiles.  The surface density profile of GCs is
  identical to that shown in Figure~\ref{radial_profile}.  The $V$
  band surface brightness profile, converted to intensity units and
  scaled to match the GC surface density profile at $r=1\arcmin$, is
  shown as the solid line.  The best-fit $R^{1/4}$ profile to the GC
  system is shown for comparison as the dashed line.  The profiles are
  shown out to a projected radius of $r=3.24\arcmin$, the radius at
  which the GC surface density profile falls to zero.
  \label{profile_comp_plot}}
\end{figure}

Figure \ref{profile_comp_plot} compares the radial surface density of
NGC~7457's GC system to the radial distribution of the galaxy light.
In the figure, the surface brightness profile derived in
$\S\ref{sfcphotometry}$ has been converted to linear (intensity) units
and scaled to the GC surface density at an arbitrary point, $r = 1
\arcmin$.  The profiles have similar shapes in the inner $\sim
1.2\arcmin$, but the GC profile remains relatively flat at larger
radii even as the galaxy light decreases exponentially. This is
consistent with the general picture of GC systems as being more
spatially extended than the parent galaxy light
\citet{ha91,as98,br06}. We caution, however, that the errors in the GC
surface density become large in the outer portions of the profile and
hence the difference between the profiles -- although suggestive -- is
not statistically significant.

\subsection{Azimuthal Distribution of the GC System}\label{ellipticity}

\begin{figure}
\plotone{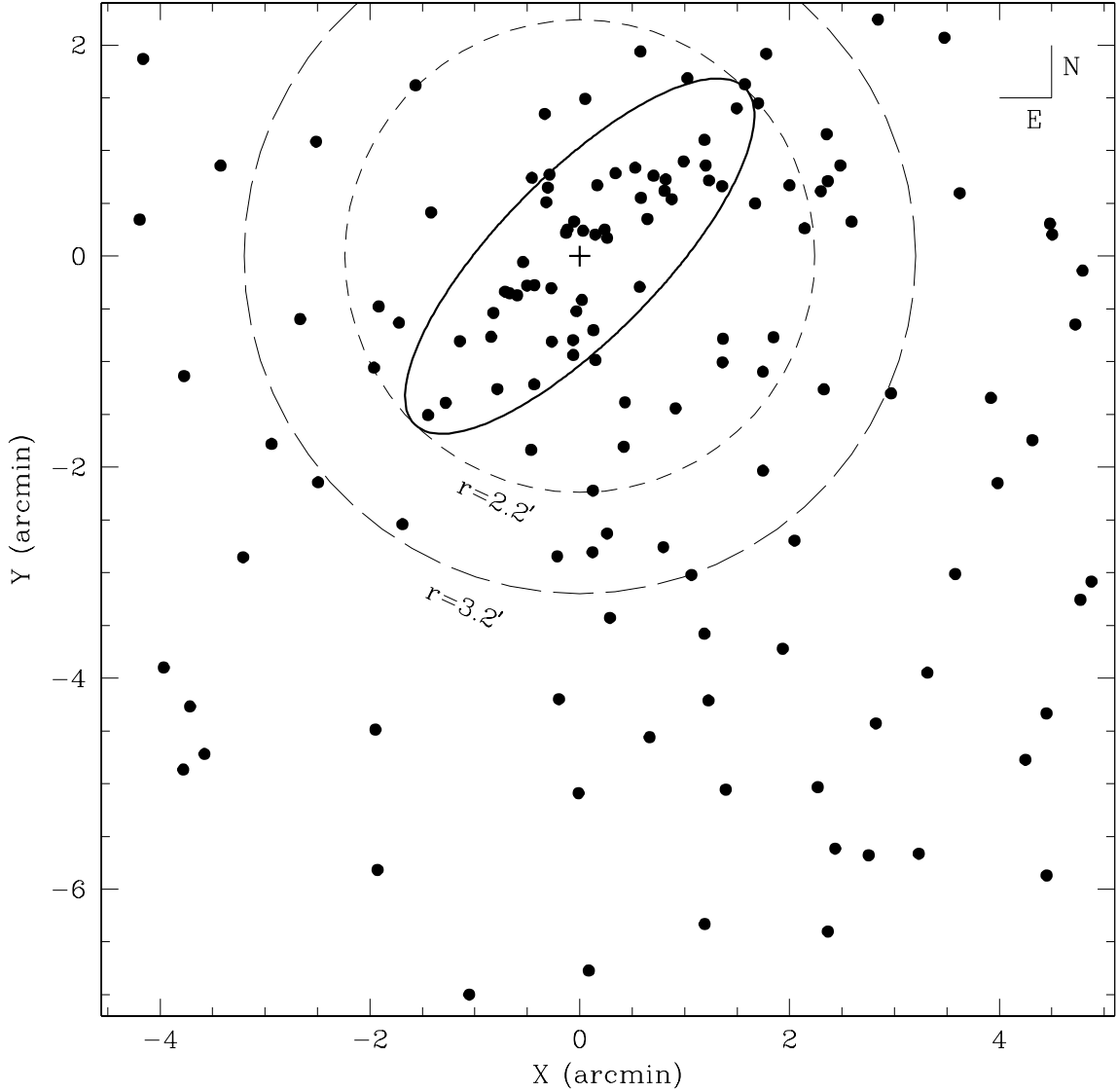}
\caption{WIYN pointing with GC candidate positions.  WIYN GC
  candidates are shown with respect to the galaxy center (plus sign).
  The radius at which the GC surface density in the final, corrected
  radial profile falls to zero is shown as the long-dashed circle.
  The radius enclosing GC candidates used in the ellipticity analysis
  $r=2.2\arcmin$ is shown as the short-dashed circle.  The ellipticity
  analysis solution (see $\S \ref{ellipticity}$) is shown as the solid
  line.
  \label{spatial_distribution_plot}}
\end{figure}

Figure~\ref{spatial_distribution_plot} shows the positions of the WIYN
GC candidates on the WIYN frame.  It is readily apparent that the GC
candidates do not show a uniform azimuthal distribution about the
galaxy center. Instead the GC system appears to be preferentially
aligned along the major axis of the galaxy in an elliptical
distribution. In order to study the azimuthal distribution of the GC
system of NGC 7457, we implemented the method of moments algorithm
\citep{tr53} to estimate the ellipticity $\epsilon$ and position angle
$\theta_{\rm p}$ of the GC system.

For our analysis we followed the methodology outlined by \citet{mc94}
and \citet{ca80}, using an iterative method to determine $\epsilon$
and $\theta_{\rm p}$.  First, the GCs within a circular aperture were
used to derive an initial unbiased estimate of $\epsilon$ and
$\theta_{\rm p}$.  This initial estimate is next used to define an
elliptical aperture and GCs within this ellipse are used to recompute
$\epsilon$ and $\theta_{\rm p}$.  For the elliptical aperture, we
scaled the semi-major axis to the radius of the initial circular
annuli to include as much of the GC population as possible in the
calculation \citep{ca80}.  The final ellipse parameters were
determined after performing these calculations to some predefined
tolerance. The method of moments also requires that the measurement
annuli lie entirely on the frame.  Because of the position of NGC~7457
on one side of the images, this requirement means we are restricted to
examining the ellipticity and position angle within the central
$2.2\arcmin$ of the galaxy (projected radius of $8.5$
kpc). Lastly, we hold the center of the GC distribution constant at
the same $X,Y$ position of the center of the galaxy light we used to
derive the GC surface density profile.

For the inner $2.2\arcmin$ of the GC system, we find an ellipticity
$\epsilon=0.66 \pm 0.14$ and position angle $\theta_{\rm
p}=-45^{\circ} \pm 10^{\circ}$ (measured east of north) determined
from 46 GC candidates. Our uncertainties were estimated via Monte
Carlo bootstrapping; comparable but slightly larger uncertainties of
$0.18$ in ellipticity and $16^{\circ}$ in position angle are found
from the simulation-derived formulae of \citet{ca80}.  In general we
find good agreement between the ellipticity and position angle of the
GC system and the galaxy isophotes, respectively.  Our derived value
of $\theta_{\rm p}=-45^{\circ} \pm 10^{\circ}$ for the GC system is
slightly lower than the average galaxy position angle of
$-52^{\circ}$, although it is consistent within the errors with the
entire range of galaxy isophote position angles.  Similarly, we
measure a larger ellipticity for the GC system ($\epsilon=0.66 \pm
0.14$) than for the galaxy light ($\bar{\epsilon} = 0.46$ from
$r=0.3\arcmin$ to $2.3\arcmin$), but this difference is only
significant at the $1.1\sigma-1.4\sigma$ level, depending on the
adopted error in the GC system ellipticity.

Because we have several masked (unmeasurable) regions on our images --
particularly the central region of the galaxy where the
galaxy-subtraction process is imperfect -- we investigated whether
these regions impart a bias into the determination of the ellipticity
and position angle.  We compared the results from two Monte Carlo
experiments of simulated GC spatial distributions: one which ignored
masked regions and one, like our observations, where the masks were
present.  For both the masked and unmasked cases we created 10,000 GC
systems and used the method of moments algorithm to derive ellipticities
and position angles. A single GC system was simulated by randomly
placing GCs in elliptical bins which matched the ellipticity, position
angle, and radial surface density profile derived for the NGC 7457 GC
system; GC positions were further restricted if the the masked regions
were being considered. The total number of GCs in each simulated
system was set to match the NGC 7457 observations. Comparing the
results from these two experiments, we found only a slight bias
towards higher ellipticities of $\Delta \epsilon \sim 0.03$ introduced
by the masked regions.  Similarly for the position angle we find only
a slight bias of $\Delta \theta_{\rm p} \sim 1^{\circ}$ towards a
larger position angle.  Although these biases are systematic and not
random uncertainties, they are well below our measurement error.

Lastly we examined the likelihood that one would measure the derived
ellipticity and position angle of the NGC 7457 GC system by chance
from a circular (i.e. azimuthally random) GC distribution.  In
general, an intrinsically circular spatial distribution analyzed via
the method of moments will systematically return a non-zero
ellipticity.  This is especially problematic when using small numbers
of objects \citep{ca80}.  So, given the total number of objects
matching our observed sample, we wished to know how often one would
measure a large ellipticity because of small number effects
\textit{even if the spatial distribution is intrinsically circular}.
To investigate this, we examined the probability of obtaining our
observed $\epsilon$ and $\theta_{\rm p}$ from a random distribution
that accounted for the masked regions.  We simulated 10,000 GC spatial
distributions with no ellipticity by placing GCs randomly in azimuth
around the galaxy center, matching the NGC 7457 observations of the
radial surface density profile and total number of GCs.  Examining the
resulting distributions of $\epsilon$ and $\theta_{\rm p}$, we find
the probability of obtaining simultaneously the measured ellipticity
($\epsilon=0.66$) and position angle ($\theta_{\rm p}=-45^{\circ}$)
within $1\sigma$ of their measured values is only $p=4.1\%$.  Hence we
can reject the hypothesis that the observed ellipticity and position
angle were observed by-chance from a azimuthally-random distribution
at the $95.9\%$ confidence level.

 \subsection{Color Distribution}\label{color}

\begin{figure}
\plotone{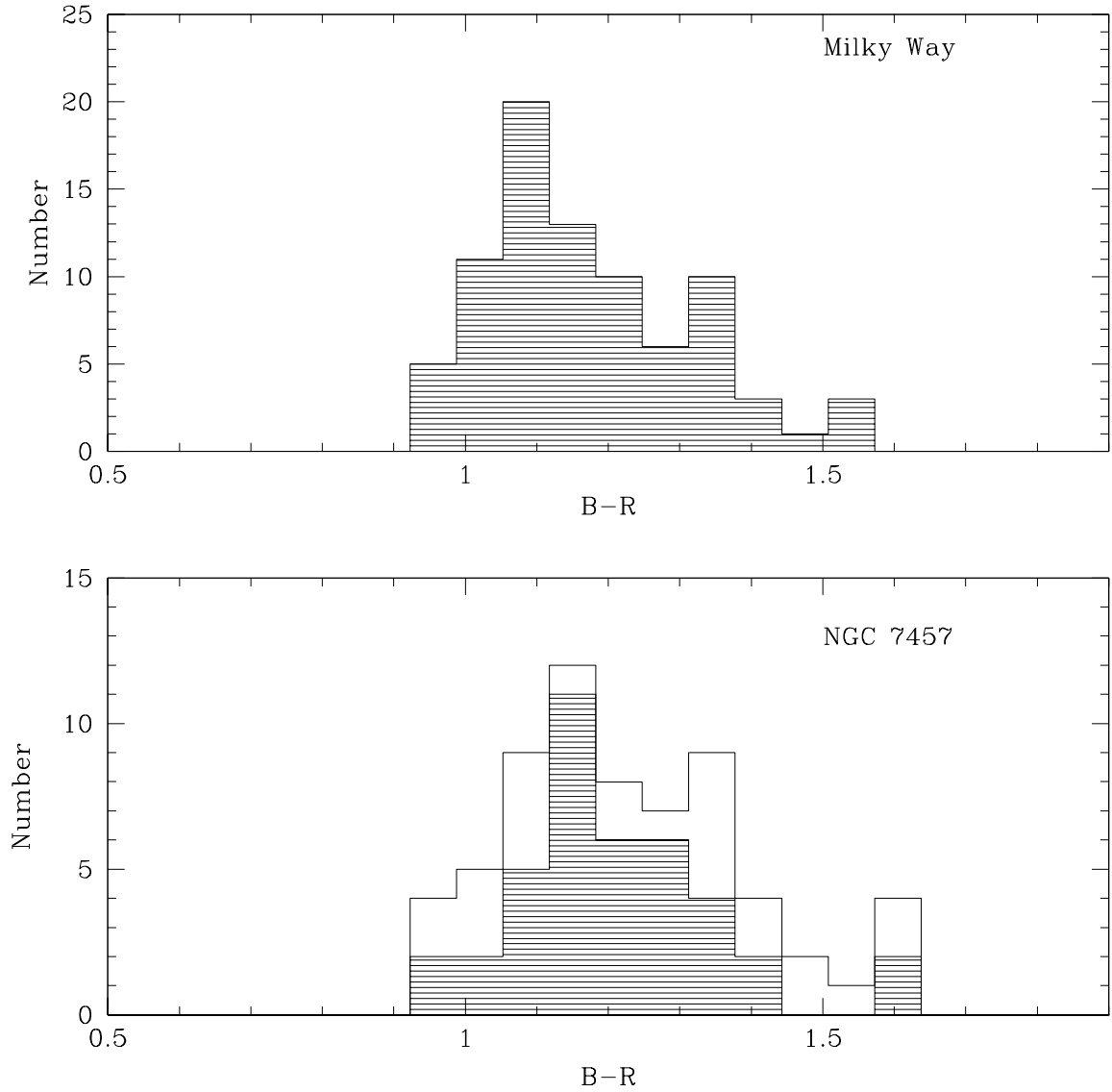}
\caption{Color distributions for the Milky Way GCs and NGC 7457 GC
  candidates.  The top panel shows the $B-R$ color distribution for 82
  Milky Way GCs with integrated $B$ and $R$ photometry from Harris
  (1996).  The bottom panel shows the $B-R$ color distribution for the
  NGC 7457 GC candidates.  The shaded regions shows the distribution
  for the radially-cut 90\% sample (40 GCs); the solid line shows the
  distribution for the full 90\% sample (65
  GCs). \label{color_distribution}}
\end{figure}

In order to investigate the relative populations of blue
(lower-metallicity) and red (higher-metallicity) GCs in NGC 7457, we
first created a subsample from our total GC candidate list that has a
magnitude completeness of $\geq 90\%$ in all three filters (see RZ01
for details and methodology).  We found 65 out of the 136 WIYN GC
candidates met this criteria and we refer to this group as the ``90\%
sample".  We can further minimize the fraction of contaminants in this
sample by imposing a radial cut on the 90\% sample at $r \leq 3.24
\arcmin$ (see $\S \ref{radial_profile}$), giving a total of 40 GC
candidates.  The $B-R$ color distribution for the 90\% sample and the
radially-constrained 90\% sample are shown in the bottom panel of
Figure~\ref{color_distribution}.  For comparison, the subset of the
Milky Way GCs with $B-R$ integrated colors (82 GCs; Harris 1996) are
shown in the top panel of Figure~\ref{color_distribution}.  A
Kolmogorov-Smirnov test rejects the hypothesis that the Milky Way GC
subset and the NGC~7457 90\% sample have the same $B-R$ distribution
at only the 89\% significance level; the K-S test does not show strong
statistical evidence to reject a common population distribution.

We examined the $B-R$ GC colors as a function of projected radius,
both for the 90\% and 90\% radially-constrained samples, to test for a
color gradient in the NGC 7457 GC system.  Linear fits to the data
showed best-fit slopes consistent with zero within the errors.  Thus
we find no evidence of a statistically significant color gradient,
although the numbers of GCs in the 90\% samples are small.

We ran the Gaussian mixture modelling (GMM) code from \citet{mu10} on
our 90\% sample.  The GMM code uses several statistics to test whether
the color distribution is better described by one or two Gaussian
functions. In addition to reproducing the results of the KMM algorithm
from \citet{as94}, the GMM code also uses the kurtosis of the
distribution and the separation of the means (relative to their
widths) as complementary tests of bimodality.  Like KMM, GMM allows
one to test both the homoscedastic (two Gaussians with the same
dispersion) and heteroscedastic (different dispersions) cases. For our
90\% sample of 65 objects, the mixture modelling results show that the
unimodal hypothesis could only be rejected at the 72\% confidence
level (heteroscedastic case).  That is, {\it we do not detect color
bimodality at a statistically significant level}.  If we also consider
the $D-$value peak separation statistic as a complementary test of
bimodality, the unimodal hypothesis could only be ruled out at the
28\% confidence level and hence this statistic also strongly argues
against bimodality.  We should stress that mixture modelling with
small numbers of objects inherently lowers the reliability of
detecting bimodality \citep{as94}.  As the number of objects
decreases, an increasingly large separation between the distribution
means is required to strongly detect bimodality.  Our results are
consistent with previous studies where bimodality was also not
detected (C99, CSB08) in NGC~7457's GC system.  Both studies performed
mixture modelling on their respective $V-I$ data and find the
distribution consistent with a single Gaussian peaking at $V-I=1.02$
(CSB08) and $V-I=1.06$ (C99; their maximum likelihood estimate). C99
also examine the $B-I$ colors and find the data consistent with a
unimodal distribution with a mean $B-I=1.91$ (their maximum likelihood
estimate). Despite the lack of an observed bimodality, both blue and
red GCs candidates are detected. C99 claimed that NGC~7457 lacks a
population of metal-poor GCs similar to the halo GCs in the Milky Way
\citep{zi93}.  In contrast, we have GC candidates (at small projected
radii, where the fraction of contaminants is lowest) with colors as
blue as the most metal-poor Galactic GCs ($B-R\sim0.93$ or [Fe/H]$\sim
-1.8$ to $-1.9$). Additionally, CSB08 have spectroscopically-confirmed
GCs in NGC 7457 with metallicities as low as ${\rm [Fe/H]}\sim -1.7$.

The broad color distribution of GCs in NGC 7457 also argues for the
presence of both metal-rich and metal-poor populations.  For our 90\%
sample, we found a mean color of $B-R=1.23$ and a distribution width
of $\sigma_{B-R}=0.17$; for comparison, the Milky Way GC system has a
mean color of $B-R=1.17$ and distribution width of $\sigma_{B-R}=0.14$
(Harris 1996; see top panel of Figure~\ref{color_distribution}).  Even
after imposing a radial cut to minimize contamination, the WIYN sample
still includes blue and red GC candidates consistent with those
observed in the Galaxy. C99 explain the observed broad distribution
and lack of bimodality as arising from the presence of the usual old
metal-rich and metal-poor populations, but with a difference in their
mean metallicities (e.g. peak separation) that is smaller than could
be detected.  CSB08 offer a different picture; they explain the broad
distribution by the presence of an intermediate-age population, which
in combination with the typical metal-rich and metal-poor populations
causes the large peak in the distribution at $V-I\sim 1.02$ ($B-R\sim
1.2$).  Spectroscopic data for a large sample of GCs in NGC~7457 would
thus be invaluable in helping to discriminate between alternative
explanations of the broad color distribution.

\subsection{Total Numbers of GCs and Global Specific Frequency}\label{numbers}

We integrated the best-fit de Vaucouleurs profile to obtain an
estimate of the total number of GCs $N_{GC}$ in NGC 7457. Although the
power law provided a slightly better fit to the surface density data
(see Table~\ref{profile_fits}), it predicts a steeply-rising surface
density at small radius (and infinite surface density at $r=0$) so is
not a physical useful model for the purposes of integration.
Integrating the de Vaucouleurs law from $r=0.08\arcmin$ (the inside
edge of the innermost radial bin) to $3.24\arcmin$ (the outer edge of
the bin where the GC surface density drops to zero within the errors)
yields 200 GCs.  We have no information about the surface density of
GCs inside $0.08\arcmin$ (308 pc for our assumed distance) so we
consider two scenarios.  The assumption that the de Vaucouleurs law
extends to $r=0$ would add another 10 GCs to the total.  Assuming
instead that the radial distribution inside $0.08\arcmin$ is flat
(with the same surface density as in the innermost bin) yields 1.5
GCs.  Averaging these results yields a final value of $N_{\rm
GC}=210\pm 30$ GCs in NGC 7457.  The error in $N_{\rm GC}$ was
calculated as described in our previous GC system studies; it includes
uncertainties in: (1) fractional coverage of the GCLF, given choices
in bin sizes and dispersion; (2) the number of GCs and contaminating
objects, assuming Poisson statistics; (3) the number of GCs in the
central, unobserved regions of the galaxy.  The uncertainties were
added in quadrature to give the total uncertainty in $N_{\rm GC}$.

For comparison, a manual integration of the GC surface density profile
was done as follows.  We assumed that the measured surface density in
each radial bin was constant across that bin, and using the calculated
area of each bin obtained a number of GCs in that bin.  Because the
overlapping WIYN and \textit{HST} bins complicate the area
calculation, we used the \textit{HST} data for $r < 2\arcmin$ and the
WIYN data for $r > 2\arcmin$.  The manual summation was performed over
the same range as the best-fit model integration ($r=0.08\arcmin$ to
$r=3.24\arcmin$) yielding a total of $208$ GCs.  If we again assume a
flat inner profile for $r<0.08\arcmin$ we estimate a total number of
GCs $N_{\rm GC}=209$, in excellent agreement with the $r^{1/4}$
profile integration.

With an estimate of the total number of GCs, we can calculate the
luminosity- and mass-normalized specific frequencies of the NGC 7457
GC system.  The luminosity-normalized specific frequency $S_N$ is
defined as total number of GCs normalized by the $V$ band luminosity
of the galaxy \citep{ha81}, or

\begin{equation}
S_N \equiv N_{\rm GC}10^{0.4(M_V+15)}.
\end{equation}

\noindent The mass-normalized specific frequency $T$ is defined as the
total number of GCs normalized by the stellar mass of the galaxy
$M_{\rm G}$ \citep{ze93}, or

\begin{equation}
T \equiv \frac{N_{\rm GC}}{M_{\rm G}/10^9 M_\odot}.
\end{equation}

\noindent Using the total absolute magnitude $M_V$ and galaxy
(stellar) mass from Table~\ref{props} with a total number of GCs
$N_{\rm GC}=210$, we find specific frequencies of $S_N=3.1 \pm 0.7$
and $T=4.8 \pm 1.1$.  The errors in $S_N$ and $T$ include the error in
$N_{\rm GC}$ as well as the uncertainty in the galaxy magnitude; the
uncertainties in these four quantities were added in quadrature to
give the total uncertainty in $S_N$ and $T$.  We assume that the
uncertainty in the galaxy magnitude is $0.2$ mag, the value of the
difference between our derived apparent magnitude and that of
\citet{fi08}.  We give our final values of $N_{\rm GC}$, $S_N$, and
$T$ in Table~\ref{sn_table}.  

C99 derived a specific frequency of $S_N=2.7\pm1.1$ from an estimated
$N_{\rm GC}=178 \pm 75$ and assumed absolute magnitude of
$M_V=-19.55$.  Thus our value of $S_N=3.1 \pm 0.7$ is consistent with
C99 within the errors. The approximately 30 ($\sim 1 \sigma$) fewer
GCs estimated by C99 compared to this study accounts for the
difference in our derived $S_N$ values given our nearly identical
assumed values of $M_V$.  The differences in $N_{\rm GC}$ are likely
due to variations in analysis techniques; C99 impose a stricter $V$
band object detection threshold as well as perform two color-color
cuts (versus our single color-color cut) with a tighter ($2\sigma$
versus our $3\sigma$) selection criteria.

\subsubsection{Mass-Normalized Specific Frequency of Metal-Poor GCs}

In order to better understand the formation of the first generation of
GCs in giant galaxies, we are interested in comparing the number of
blue, metal-poor GCs in this galaxy to the number for other galaxies
\citep{rh05}.  Because the mixture modelling results argue against
color bimodality, we do not adopt the resulting red and blue fractions
of GCs from these tests.  Instead, we take a very simple approach and
estimate the number of blue and red clusters using the
empirically-determined color separation between the blue and red
populations for elliptical galaxies of $B-R\sim 1.23$ (RZ01, RZ04,
Kundu \& Whitmore 2001a).  For the 90\% sample, we find that 55\% of
the GC candidates have a $B-R$ color bluer than 1.23. If we impose a
radial cut on the 90\% sample ($r\leq 3.24\arcmin$; see $\S
\ref{radial_profile}$) in order to minimize contamination, we find
that 60\% of GC candidates have a $B-R$ color bluer than 1.23.  If we
take the fraction of blue GCs in our total sample as the average of
these two estimates, we find the mass-normalized number of blue GCs of
$T_{\rm blue}=2.8 \pm 0.6$.  Compared to the other galaxies in our
wide-field survey (see R10), this value of $T_{\rm blue}$ is higher
than other galaxies of similar mass.  By morphological type, the
weighted mean values of $T_{\rm blue}$ for the spirals and ellipticals
in our wide-field survey are $0.82\pm0.10$ (9 spirals, including the
Milky Way and M31; R10)and $1.4\pm0.1$ (7 ellipticals; R10),
respectively.  Although NGC 7457 appears to have quite a high value of
$T_{\rm blue}$ compared to our survey results, NGC 7457 is currently
the lowest mass galaxy in our sample and therefore few galaxies are
available for direct, systematic comparision. The addition to our
survey of galaxies in the low mass regime ($\log(M/M_\odot)<10.8$)
will be critical to understanding the value of $T_{\rm blue}$ for NGC
7457 and other low mass galaxies.

\section{Summary and Conclusions}\label{conclusions}

This study investigated the global properties of the GC system of the
field S0 galaxy NGC 7457, combining wide-field WIYN Minimosaic imaging
data with archival and published {\it HST}/WFPC2 data.  Here we
summarize our main results:

\begin{enumerate}

\item We performed surface photometry on the WIYN $V$ image of
NGC~7457 and measure a total magnitude of $V_T^0$ $=$ 11.07. The light
profile for the galaxy follows a de~Vaucouleurs law in the inner $\sim
0.3\arcmin$ and declines exponentially from $\sim 0.3\arcmin$ to
$2.3\arcmin$, the limit of our photometry.

\item We constructed a radial surface density profile for the GC
system using both the WIYN and HST data.  The surface density of the
system goes from $74~ {\rm arcmin}^{-2}$ at $0.4\arcmin$ (1.4 kpc) to
zero (within the errors) at $3.24\arcmin$ (12.5 kpc). The GC surface
density and galaxy surface brightness profiles are similar in the
inner $1 \arcmin$ (3.9 kpc). Although the GC profile appears to
flatten at larger radii relative to the galaxy light profile, the
uncertainties in the surface density in the outer portions of the
profile are significant.

\item The GC radial surface density profile was fit with both de
Vaucouleurs and power laws.  From the integration of the best-fit de
Vaucouleurs profile, we derive a total number of GCs of $N_{\rm GC} =
210 \pm 30$.  Using our galaxy photometry, we found a $V$-band
normalized specific frequency of $S_N = 3.1 \pm 0.7$, a
mass-normalized specific frequency of $T = 4.8 \pm 1.1$, and a
mass-normalized number of blue (metal-poor) GCs of $T_{\rm blue} = 2.8
\pm 0.6.$

\item The GCs in NGC 7457 are distributed in an elliptical
configuration, with the major axis of the GC system aligned with the
galaxy's major axis. We measure a GC system ellipticity $\epsilon =
0.66 \pm 0.14$, which is consistent with the average ellipticity of
the galaxy.

\item We find that the B-R color distribution of NGC~7457's GC system
is not significantly bimodal. However, the color distribution appears
similar to that of the Milky Way GC system, with both very blue and
very red GC candidates close to the galaxy center, where contamination
from non-GCs should be minimized in our sample.  Lastly, we find no
evidence for a color gradient in NGC~7457's GC system.

\end{enumerate}

We can explore these results in light of various formation and
evolution models for S0 galaxies. The typical formation scenario for
lenticular galaxies involves the transformation of S0s from spiral
galaxies as star formation is either suppressed or altogether halted
due to the removal of gas from the galaxy.  Observational support for
such a transformation scenario comes from tracing the decreasing ratio
of spiral to S0 galaxies in galaxy clusters with decreasing cluster
redshift \citep{dr97}.  Naturally then, mechanisms for this
spiral-lenticular evolution have generally focused on processes for
disk gas removal in a cluster environment.  \citet{bo06} give an
overview of such mechanisms, including tidal interactions,
ram-pressure stripping, galaxy harassment, viscous stripping, thermal
evaporation, and galaxy starvation/strangulation.  However, because
NGC 7457 is located in the field, we are particularly interested in S0
galaxy formation scenarios that do not rely on a cluster environment
for gas removal or star-formation suppression.

The luminosity-normalized specific frequency of GCs $S_N$ has recently
been proposed as a means of tracing ``galaxy fading'' in lenticulars
\citep{ar06,ba07}.  If S0 galaxies form from gas removal in spirals
then, as star formation declines and the stellar populations age, the
total $V$-band luminosity of the galaxy will decrease and the global
color of the galaxy should become more red.  If in addition the total
number of GCs in the galaxy remains constant (a key assumption which
ignores GC destruction mechanisms), the decreasing $V$-band magnitude
should lead to an increasing $S_N$ as the system ages.  Such a
correlation between $S_N$ and galaxy color was found by \citet{ar06}
using the global galaxy colors from the photometry of \citet{po94} and
\citet{pr98} with the ``local'' $S_N$ values from the {\it HST}/WFPC2
sample of lenticulars from \citet{ku01b}. Here the local specific
frequency is defined as the specific frequency within the {\it
HST}/WFPC2 field-of-view, in contrast to a global value derived from
observations like ours, where the full radial extent of the GC system
and galaxy light have been observed.

To compare our results for NGC 7457 to this trend, we estimated a
local specific frequency of $S_N \sim 2.5$ by only considering the GC
radial profile and integrated galaxy light in the central
$2.2\arcmin$.  From \citet{pr98}, the global galaxy colors for NGC
7457 are $U-B=0.36\pm0.2$, $B-V=0.90\pm0.01$, and $V-R=0.56\pm0.05$. We
find that NGC 7457 is not consistent with the observed color-$S_N$
trend of \citet{ar06}, particularly in the $(U-B)-S_N$ plane where the
correlation has the least amount of scatter (see their Figure~2).  NGC
7457 appears to be either too blue for its $S_N$ or has a much larger
$S_N$ for its observed color.  Although NGC~7457 shows better
agreement in the $(B-V)-S_N$ and $(V-R)-S_N$ planes, the overall
scatter in the trends are significant.  While an inconsistency with
the global color-$S_N$ trend may suggest that fading has been largely
unimportant in the evolution of NGC 7457, it should be noted that this
relation was derived from a sample comprised almost entirely of
cluster or group S0 galaxies.  Thus the galaxy color-$S_N$ trend
likely includes an environmental bias and should be explored with a
sample of lenticulars which cover a wider range of environments.

As an alternative model of S0 galaxy formation, \citet{be98} and
\citet{bo05} have shown that mergers of unequal-mass, gas-rich spirals
(mass ratio $\sim 0.3$) can produce remnants with the observed
properties of S0s. In the \citet{be98} simulations, vertical dynamical
heating in the disk of the more massive galaxy creates the thickened
disk component often observed in S0 galaxies.  The tidal interations
and subsequent merging of the galaxies remove the majority of the gas
in two moderate bursts of star formation, resulting in a central
concentration of new stars.  Similarly, major mergers have been
proposed as a formation mechanism for new populations of GCs
\citep{sw87,as92}. The larger GC systems of ellipticals relative to
spirals (as traced by specific frequencies) may be due at least in
part to additional GCs formed during the mergers. Therefore a merger,
if it were gas-rich and created GCs, might help explain the relatively
large specific frequencies of NGC 7457 ($S_N = 3.1 \pm 0.7$, $T = 4.8
\pm 1.1$) for its mass. For comparison, we find average specific
frequencies for ellipticals of $S_N = 2.6 \pm 0.4$, $T = 3.1 \pm 0.5$
(6 galaxies; R10) versus $S_N = 0.8 \pm 0.1$, $T = 1.5 \pm 0.2$ for
spirals (9 galaxies including M31 and the Galaxy; R10).  By total
numbers of GCs, however, NGC 7457 is more consistent ($N_{\rm GC} =210
\pm 30$) with the average value for spirals ($190 \pm 40$; 9 galaxies,
including M31 and the Galaxy). Determining the global properties of GC
systems for a larger sample of S0s will be critical in refining the
comparison across morphological types.

Additional evidence for a merger origin for NGC 7457 comes from the
discovery of a counter-rotating galaxy core \citep{si02}.  Such
kinematically distinct cores are thought to be signatures of merger
events (see Forbes et al. 1996 and references therein) and can arise
in simulations of unequal mass spiral-spiral mergers \citep{ba98}.
\citet{si02} studied NGC 7457 using integral field unit spectroscopy
and derived a two dimensional line-of-sight velocity map of the
galaxy's central region. These data also provided absorption line
equivalent widths; using Lick indices \citep{wo94}, ages and
metallicities of the stellar populations were explored.  In addition
to observing a counter-rotating core, \citet{si02} found that the core
was significantly younger (age $\sim 2-2.5$ Gyr) and more metal-rich
(approximately solar) than the surrounding bulge region (age $\sim
5-7$ Gyr; approximately one-half solar metallicity).  CSB08 provided
circumstantial evidence for a GC population with an age comparable to
that of the core ($\sim 2-3$ Gyr), noting that this GC population
could have been formed in the same star formation event that resulted
in the young, chemically-distinct core.  If such an intermediate age
GC population is confirmed, this could be additional evidence for an
increased number of GCs resulting from star formation in a merger
event.

The evidence from surface photometry of a classical bulge in NGC 7457
(Fisher \& Drory 2008, 2010; see $\S \ref{sfcphotometry}$) is also
suggestive of a merger history in this galaxy.  Classical bulges are
thought to arise as a result of merger events, while pseudobulge
formation is associated with internal dynamic processes such as bars,
ovals, and spiral structure \citep{ko04,co09}.  As noted by CSB08, the
large number of metal-rich GCs in NGC 7457 would also be consistent
with a merger-built bulge, assuming such mergers were sufficiently
gas-rich and included GC formation.  However, it is not clear why the
central velocity dispersion of this galaxy ($\sim 60~\rm{km s}^{-1}$;
Sil'chenko et al. 2002, Cherepashchuk et al. 2010) is low like a
pseudobulge if the bulge was formed from mergers; merger events should
give rise to a dynamically-hot bulge (see Fisher \& Drory 2010 and
references therein).  Furthermore, both the estimated total number of
GCs in NGC~7457 and the GC $B-R$ color distribution are similar to the
Milky Way, which has a pseudobulge \citep{ko04} and may only have a
small merger-built classical bulge component \citep{sh10}. Additional
studies of the core and bulge regions in NGC 7457 will be necessary to
further explore the role of merger or secular processes.

Further insights into formation scenarios could be gained from a
large-sample spectroscopic study of GCs in NGC 7457.  Although we have
detected a very elliptical GC spatial distribution, without kinematic
data it is difficult to interpret this observation.  An inclined,
disky population of GCs could possibly reproduce the observed
elliptical distribution.  The dynamics of GC systems in E/S0s formed
in spiral-spiral mergers has been explored by \citet{be05}; kinematic
data on GCs in NGC 7457 would give an additional means of
investigating a possible merger history in this galaxy.  Thus future
spectroscopic studies of GCs in NGC 7457 could provide key data for
understanding the formation and evolution of this galaxy, both in
determining GC ages and exploring the kinematics of the GC system.

\acknowledgments

This research was supported by an NSF Faculty Early Career Development
(CAREER) award (AST-0847109) to K.L.R.  We are grateful to Mike Young
for acquiring the data for this study and to the staff of the WIYN
Observatory and Kitt Peak National Observatory for their assistance
during the observing run.  J.R.H. would like to thank Steven
Janowiecki and Liese van Zee for useful conversations during the
course of this work. We would like to thank the anonymous referee for
his/her useful comments which have improved the quality of the
paper. This research has made use of the NASA/IPAC Extragalactic
Database (NED) which is operated by the Jet Propulsion Laboratory,
California Institute of Technology, under contract with the National
Aeronautics and Space Administration.



\clearpage

\clearpage

\begin{table}
\begin{center}
\caption{Properties of NGC 7457.\label{props}}
\begin{tabular}{lcl}
\tableline\tableline
Parameter & Value & Reference\\
\tableline
Type              & S01            & \citet{sa61} \\
$V_T^0$           & 11.07          & This study \\
$m-M$             & $30.61\pm0.21$ & \citet{to01} \\
$M_V$             & -19.54         & $V_T^0$ and $(m-M)$ \\
$A_V$             & 0.168          & \citet{sc98} \\
Distance          & 13.2 Mpc       & \citet{to01} \\
$v_{\rm helio}$   & 812 ${\rm km~ s}^{-1}$ & \citet{si97} \\
Inclination       & $59^{\circ}$   & This study\\
$\log(M/M_\odot)$ & 10.6          & $(M/L)_V$ and $M_V$ \\
\tableline
\end{tabular}
\end{center}
\end{table}

\begin{deluxetable}{lcccc}
\tablecaption{Surface Photometry for NGC 7457.\label{sfc_phot_table}}
\tablehead{\colhead{SMA (arcsec)} & \colhead{$V\pm{\sigma}_V$} &
\colhead{$\mu_V ~({\rm mag ~arcsec}^{-2})$} & \colhead{$\epsilon \pm
{\sigma}_\epsilon$} & \colhead{${\rm PA} \pm {\sigma}_{\rm PA}$
(degrees)}
}
\startdata
  2.10 & $17.460 \pm 0.013$ & 18.103 & $0.2720 \pm 0.0000$ & $-53.91 \pm 0.00$ \\
  2.31 & $16.637 \pm 0.013$ & 18.204 & $0.2720 \pm 0.0014$ & $-53.91 \pm 0.21$ \\
  2.54 & $16.175 \pm 0.013$ & 18.305 & $0.2844 \pm 0.0014$ & $-53.44 \pm 0.21$ \\
  2.79 & $15.849 \pm 0.013$ & 18.408 & $0.2975 \pm 0.0015$ & $-52.43 \pm 0.26$ \\
  3.07 & $15.562 \pm 0.013$ & 18.509 & $0.3098 \pm 0.0012$ & $-52.38 \pm 0.20$ \\
  3.38 & $15.325 \pm 0.013$ & 18.608 & $0.3197 \pm 0.0048$ & $-53.23 \pm 0.72$ \\
  3.72 & $15.108 \pm 0.013$ & 18.716 & $0.3276 \pm 0.0014$ & $-52.25 \pm 0.20$ \\
  4.09 & $14.925 \pm 0.013$ & 18.810 & $0.3424 \pm 0.0067$ & $-52.52 \pm 0.82$ \\
  4.50 & $14.732 \pm 0.013$ & 18.928 & $0.3424 \pm 0.0034$ & $-52.52 \pm 0.37$ \\
  4.94 & $14.566 \pm 0.013$ & 19.034 & $0.3493 \pm 0.0024$ & $-52.64 \pm 0.21$ \\
  5.44 & $14.408 \pm 0.013$ & 19.145 & $0.3545 \pm 0.0007$ & $-52.22 \pm 0.06$ \\
  5.98 & $14.263 \pm 0.013$ & 19.254 & $0.3605 \pm 0.0006$ & $-51.96 \pm 0.05$ \\
  6.58 & $14.122 \pm 0.013$ & 19.363 & $0.3652 \pm 0.0007$ & $-51.77 \pm 0.06$ \\
  7.24 & $13.990 \pm 0.013$ & 19.469 & $0.3719 \pm 0.0009$ & $-51.59 \pm 0.08$ \\
  7.96 & $13.865 \pm 0.013$ & 19.571 & $0.3804 \pm 0.0009$ & $-51.44 \pm 0.09$ \\
  8.76 & $13.741 \pm 0.013$ & 19.669 & $0.3877 \pm 0.0016$ & $-51.20 \pm 0.14$ \\
  9.64 & $13.621 \pm 0.013$ & 19.759 & $0.3961 \pm 0.0017$ & $-50.33 \pm 0.16$ \\
 10.60 & $13.483 \pm 0.013$ & 19.853 & $0.3862 \pm 0.0032$ & $-48.85 \pm 0.29$ \\
 11.66 & $13.388 \pm 0.013$ & 19.927 & $0.4145 \pm 0.0021$ & $-48.53 \pm 0.18$ \\
 12.83 & $13.289 \pm 0.013$ & 20.006 & $0.4369 \pm 0.0011$ & $-48.40 \pm 0.09$ \\
 14.11 & $13.170 \pm 0.013$ & 20.105 & $0.4405 \pm 0.0010$ & $-48.27 \pm 0.08$ \\
 15.52 & $13.043 \pm 0.013$ & 20.224 & $0.4347 \pm 0.0011$ & $-48.26 \pm 0.08$ \\
 17.07 & $12.928 \pm 0.013$ & 20.334 & $0.4366 \pm 0.0009$ & $-48.19 \pm 0.08$ \\
 18.78 & $12.811 \pm 0.013$ & 20.452 & $0.4349 \pm 0.0009$ & $-48.43 \pm 0.07$ \\
 20.66 & $12.695 \pm 0.013$ & 20.577 & $0.4317 \pm 0.0010$ & $-49.08 \pm 0.08$ \\
 22.72 & $12.588 \pm 0.013$ & 20.701 & $0.4339 \pm 0.0012$ & $-50.20 \pm 0.10$ \\
 24.99 & $12.488 \pm 0.013$ & 20.828 & $0.4406 \pm 0.0011$ & $-51.03 \pm 0.09$ \\
 27.49 & $12.393 \pm 0.013$ & 20.952 & $0.4493 \pm 0.0007$ & $-51.56 \pm 0.06$ \\
 30.24 & $12.303 \pm 0.013$ & 21.079 & $0.4595 \pm 0.0007$ & $-52.25 \pm 0.06$ \\
 33.27 & $12.217 \pm 0.013$ & 21.200 & $0.4729 \pm 0.0007$ & $-52.72 \pm 0.06$ \\
 36.59 & $12.124 \pm 0.013$ & 21.355 & $0.4752 \pm 0.0007$ & $-53.54 \pm 0.06$ \\
 40.25 & $12.032 \pm 0.013$ & 21.530 & $0.4729 \pm 0.0005$ & $-53.84 \pm 0.05$ \\
 44.28 & $11.946 \pm 0.013$ & 21.693 & $0.4737 \pm 0.0005$ & $-54.48 \pm 0.04$ \\
 48.70 & $11.851 \pm 0.013$ & 21.921 & $0.4561 \pm 0.0006$ & $-54.44 \pm 0.05$ \\
 53.57 & $11.779 \pm 0.013$ & 22.098 & $0.4627 \pm 0.0007$ & $-53.24 \pm 0.06$ \\
 58.93 & $11.718 \pm 0.013$ & 22.235 & $0.4807 \pm 0.0007$ & $-53.24 \pm 0.06$ \\
 64.83 & $11.653 \pm 0.013$ & 22.412 & $0.4886 \pm 0.0006$ & $-53.98 \pm 0.05$ \\
 71.31 & $11.585 \pm 0.013$ & 22.644 & $0.4870 \pm 0.0006$ & $-54.18 \pm 0.05$ \\
 78.44 & $11.520 \pm 0.013$ & 22.925 & $0.4797 \pm 0.0006$ & $-53.78 \pm 0.05$ \\
 86.28 & $11.465 \pm 0.013$ & 23.224 & $0.4778 \pm 0.0007$ & $-53.56 \pm 0.06$ \\
 94.91 & $11.417 \pm 0.013$ & 23.557 & $0.4731 \pm 0.0014$ & $-53.87 \pm 0.12$ \\
104.40 & $11.373 \pm 0.013$ & 23.960 & $0.4613 \pm 0.0013$ & $-53.45 \pm 0.11$ \\
114.84 & $11.333 \pm 0.014$ & 24.496 & $0.4315 \pm 0.0014$ & $-54.04 \pm 0.13$ \\
126.33 & $11.305 \pm 0.014$ & 25.007 & $0.4108 \pm 0.0042$ & $-54.04 \pm 0.39$ \\
138.96 & $11.285 \pm 0.014$ & 25.498 & $0.3979 \pm 0.0035$ & $-55.61 \pm 0.32$ \\
\enddata
\tablecomments{SMA is the semi-major axes of the
  elliptical isophotes.  The values of $V$ are the total magnitude of the
  galaxy interior to SMA.  The values of PA are the position angle of
  the ellipse isophotes measured in degrees east of north.
  }
\end{deluxetable}

\begin{deluxetable}{lccc}
\tablecaption{Corrected Radial Surface Density Profile of the GC System of
  NGC 7457\label{radial_profile_table}}
\tablehead{\colhead{Radius} & \colhead{Surface Density} &
  \colhead{Fractional Coverage} & \colhead{Source}}
\startdata
0.4 &  $73.77 \pm 15.38$ & 0.66  & $HST$ \\
0.6 &  $23.20 \pm  5.46$ & 0.78  & WIYN \\
0.9 &  $21.35 \pm  5.34$ & 0.63  & $HST$ \\
1.0 &   $7.73 \pm  2.32$ & 0.93  & WIYN \\
1.3 &  $ 6.34 \pm  2.59$ & 0.50  & $HST$ \\
1.5 &  $ 4.60 \pm  1.74$ & 0.86  & WIYN \\
1.7 &  $ 4.08 \pm  4.08$ & 0.09  & $HST$ \\
2.0 &   $3.62 \pm  1.50$ & 0.89  & WIYN \\
2.5 &  $ 1.97 \pm  1.30$ & 0.88  & WIYN \\
3.0 &  $ 0.26 \pm  1.15$ & 0.74  & WIYN \\
3.5 &  $ 0.72 \pm  1.18$ & 0.70  & WIYN \\
4.0 &  $ 0.41 \pm  1.15$ & 0.68  & WIYN \\
4.5 &  $ 0.70 \pm  1.17$ & 0.62  & WIYN \\
5.0 &  $ 0.19 \pm  1.14$ & 0.48  & WIYN \\
5.5 &  $ 0.47 \pm  1.19$ & 0.34  & WIYN \\
6.0 &  $ 1.09 \pm  1.27$ & 0.29  & WIYN \\
6.5 &  $ 0.36 \pm  1.20$ & 0.25  & WIYN \\
7.0 &  $ 0.07 \pm  1.16$ & 0.23  & WIYN \\
7.5 &  $-0.32 \pm  1.15$ & 0.14  & WIYN \\
\enddata
\tablecomments{Negative surface densities can occur due to the
 application of the contamination correction.}
\end{deluxetable}

\begin{table}
\begin{center}
\caption{GC Radial Surface Density Profile Fit
Parameters.\label{profile_fits}}
\begin{tabular}{lccc}
\tableline\tableline
Profile Fit & $a_0$  & $a_1$ & $\chi^2/\nu$ \\
\tableline
Power Law & $1.03 \pm 0.04$ & $-1.82 \pm 0.13$ & 0.70 \\
de Vaucouleurs Law & $4.09 \pm 0.22$ & $-3.01 \pm 0.22$ & 0.92 \\
\tableline
\end{tabular}
\end{center}
\end{table}

\begin{table}
\begin{center}
\caption{Total Number and Specific Frequencies for the GC System of
  NGC 7457.\label{sn_table}}
\begin{tabular}{lc}
\tableline\tableline
Parameter & Value \\
\tableline
$N_{\rm GC}$ & $210 \pm 30$ \\
$S_N$        & $3.1 \pm 0.7$ \\
$T$          & $4.8 \pm 1.1$ \\
$T_{\rm blue}$& $2.8 \pm 0.6$ \\
\tableline
\end{tabular}
\end{center}
\end{table}

\end{document}